\documentclass[preprint,prb]{revtex4-1} 

\usepackage{amsmath}  
\usepackage{amsfonts} 
\usepackage{graphicx} 
\usepackage{multirow}
\usepackage{color}
\usepackage[table]{xcolor}
\usepackage{scrextend}

\begin{document}

\title{Introducing Quantum Entanglement to First-Year Students: Resolving the Trilemma}

\author{W.M. Stuckey}
\affiliation{Department of Physics, Elizabethtown College, Elizabethtown, PA 17022}
\author{Timothy McDevitt}
\affiliation{Department of Mathematical Sciences, Elizabethtown College, Elizabethtown, PA 17022}
\author{Michael Silberstein}
\affiliation{Department of Philosophy, Elizabethtown College, Elizabethtown, PA 17022}
\affiliation{Department of Philosophy, University of Maryland, College Park, MD 20742}
\date{16 March 2024}

\begin{abstract}
While quantum mechanics (QM) is covered at length in introductory physics textbooks, the concept of quantum entanglement is typically not covered at all, despite its importance in the rapidly growing area of quantum information science and its extensive experimental confirmation. Thus, physics educators are left to their own devices as to how to introduce this important concept. Regardless of how a physics educator chooses to introduce quantum entanglement, they face a trilemma involving its mysterious Bell-inequality-violating correlations. They can compromise on the the completeness of their introduction and simply choose not to share that fact, totally ignoring the 2022 Nobel Prize in Physics. They can frustrate their more curious students by introducing the mystery and simply telling them that the QM formalism with its associated (equally mysterious) conservation law maps beautifully to the experiments, so there is nothing else that needs to be said. Or, they can compromise the rigor of their presentation and attempt to resolve the mystery by venturing into the metaphysical quagmire of competing QM interpretations. Herein, we resolve this trilemma in precisely the same way that Einstein resolved the mysteries of time dilation and length contraction that existed in the late nineteenth century. That is, we resort to ``principle'' explanation based on the mathematical consequences of ``empirically discovered'' facts. Indeed, our principle account of quantum entanglement is even based on the same principle Einstein used, i.e., the relativity principle or ``no preferred reference frame.'' Thus, this principle resolution of the trilemma is as complete, satisfying, analytically rigorous, and accessible as the standard introduction of special relativity for first-year physics students. 
\end{abstract}

\maketitle


\section{Introduction} \label{SecIntro}

\noindent A review of some introductory physics textbooks\cite{walker,giancoli,knight,tipler,serway} reveals what Ross recently pointed out about textbooks on modern physics\cite{ross2020}, ``None of the popular texts include topics on Dirac notation, quantum entanglement or quantum computing-quantum information.'' The one exception\cite{young} has a section on quantum entanglement and does mention quantum computing, but says nothing about quantum information science. This is a serious omission because the concept of quantum entanglement is central to the important new field of quantum information science, as recognized in the 2022 Nobel Prize in Physics awarded\cite{nobel2022} ``for experiments with entangled photons, establishing the violation of Bell inequalities and pioneering quantum information science.'' Bub writes\cite{Bub2020}:
\begin{quote}
    A pair of quantum systems in an entangled state can be used as a quantum information channel to perform computational and cryptographic tasks that are impossible for classical systems. ... any attempt by Eve to measure the quantum systems in the entangled state shared by Alice and Bob will destroy the entangled state. Alice and Bob can detect this by checking a Bell inequality.
\end{quote}
This practical reason alone justifies an introduction to quantum entanglement for students taking the introductory physics sequence for scientists and engineers (hereafter ``first-year students''). 

In their defense, introductory physics textbooks can hardly be expected to cover this material since the Topical Group of Quantum Information\cite{DQI} was only officially established as a Division of the American Physical Society in 2017 and introductory textbooks should restrict their coverage to the most well-established material. It is really up to physics educators to augment their course content with exciting new material. As Wheeler once said\cite[p. 213]{wheeler1990}: 
\begin{quote}
    To tell their students something both new and true, something that will grip them with its power and surprise, is the time-honored obligation of teacher-researchers. 
\end{quote}
Indeed, there was an entire session (A29) in the 2021 APS March Meeting\cite{APS} dedicated to ``Quantum Information Education'' and the March 2021 edition of \textit{Physics Today} contains a ``Special Focus on Quantum Information.'' In his editorial, ``Quantum information is exciting and important,'' Charles Day mentions\cite[p. 8]{PhysicsToday2021} ``entanglement-based clocks of unprecedented precision'' and points out that a ``76-qubit device based on entangled photons solved a sampling problem $10^{14}$ times faster than a classical device could.'' 

While there are no serious hurdles for introducing quantum entanglement to first-year students, there is a trilemma concerning what to say about its mysterious Bell-inequality-violating correlations, which has been called\cite{aczel} ``The Greatest Mystery in Physics.'' The mystery of quantum entanglement per se isn't the problem, it is easy to introduce conceptually to first-year students using any number of actual or imagined experiments, e.g., the GHZ experiment\cite{merminRevisited}, Hardy's experiment\cite{merminRefined}, Zeilinger's delayed choice experiment\cite{zeilinger}, or Kim et al.'s delayed choice quantum eraser experiment\cite{kim}. Indeed, Dehlinger \& Mitchell even published an experiment\cite{dehlinger} revealing the violations of a Bell inequality\cite{bell} that is suitable for the undergraduate physics lab, to include their data that students may analyze. The problem is how to \textit{resolve} that mystery for the students without being drawn into contentious metaphysical arguments concerning\cite{becker} ``What is Real?'' Here is the trilemma concerning what the physics educator can do about the fact that there is a mystery associated with quantum entanglement. They can:
\begin{enumerate}
\item Compromise on the completeness of their introduction and simply choose not to share that fact. The problem here is that most of today’s students will know something very interesting has been omitted, since the 2022 Nobel Prize in Physics recognized ``the violation of Bell inequalities.''
\item Introduce the mystery and simply tell the students that since the formalism of quantum mechanics (QM) with its associated conservation law maps so beautifully to the confirming experiments, there is nothing else that needs to be said. This attitude of ``Shut up and calculate!'' is the ``Copenhagen interpretation of QM'' according to Mermin\cite{merminShutUp}. It was by far the favorite interpretation of QM at 42\% in a 2011 poll by Schlosshauer et al\cite{schlosshauer2013} (second place was ``information-theoretic'' at 24\%). The problem here is, as we will see, the type of conservation at work in the Bell states is very different than in classical physics, i.e., it holds only \textit{on average} when Alice and Bob are making different measurements. Thus, `average-only' conservation is really \textit{an articulation of the mystery, not a resolution of the mystery}. Consequently, ``Shut up and calculate!'' won't satisfy the students' curiosity today any more than it did Mermin's when he was a student. Again, the students will certainly feel that something important has been omitted.
\item Introduce the mystery and attempt to resolve it via some metaphysical interpretation(s) of QM. The problem here is that the educator has now ventured into the metaphysical quagmire of competing QM interpretations. For example, Drummond's 2019 overview of QM interpretations, ``Understanding quantum mechanics: a review and synthesis in precise language,'' is 48 pages long with 570 references\cite{drummond2019}. Since no experiment can distinguish between metaphysical models and there is no consensus metaphysical model to present, the educator has seriously compromised the rigor of their presentation.
\end{enumerate}
This trilemma resides in the desire for a \textit{constructive} account of quantum entanglement, so we resolve it herein by pivoting to a \textit{principle} account of quantum entanglement, just as Einstein did to resolve the mysteries of time dilation and length contraction associated with the\cite{brown2003} ``FitzGerald-Lorentz contraction hypothesis ... a cornerstone of the `kinematic' component of the special relativity (SR).'' According to Einstein, a constructive theory is based on dynamical laws and/or mechanistic causal processes (causal mechanisms) while a principle theory is based on an empirically discovered fact\cite{einstein1919}. He used the kinetic theory of gases as an example of a constructive theory and thermodynamics as an example of a principle theory where the empirically discovered fact at its foundation is ``perpetual motion machines are impossible.'' SR is also a principle theory based on the empirically discovered fact that everyone measures the same value for the speed of light $c$, regardless of their uniform relative motions (light postulate).

Fuchs writes\cite{fuchs2002}:
\begin{quote}
Where present-day quantum-foundation studies have stagnated in the stream of history is not so unlike where the physics of length contraction and time dilation stood before Einstein's 1905 paper on special relativity.
\end{quote}
In other words, SR provides an historical precedent for dealing with the trilemma of quantum entanglement\cite{moylan2022}:
\begin{quote}
To put things into an historical perspective, we recall that at the end of the nineteenth century, physics was in a terrible state of confusion. Maxwell's equations were not preserved under the Galilean transformations, and most of the Maxwellian physicists of the time were ready to abandon the relativity of motion principle (Refs. 10 and 11). They adopted a distinguished frame of reference, the rest frame of the ``luminiferous aether,'' as the medium in which electromagnetic waves propagate and in which Maxwell's equations and the Lorentz force law have their usual forms. In effect, they were ready to uproot Copernicus and reinstate a new form of geocentrism.
\end{quote}
Even ``Einstein was willing to sacrifice the greatest success of 19th century physics, Maxwell’s theory, seeking to replace it by one conforming to an emission theory of light, as the classical, Galilean kinematics demanded'' before realizing that such an emission theory would not work\cite[p. 38]{norton2004}. In their introduction of SR, Serway \& Jewett write\cite[p. 1016]{serway}:
\begin{quote}
    To resolve this contradiction in theories, we must conclude that either (1) the laws of electricity and magnetism are not the same in all inertial frames or (2) the Galilean velocity transformation equation is incorrect. If we assume the first alternative, a preferred reference frame in which the speed of light has the value \textit{c} must exist and the measured speed must be greater or less than this value in any other reference frame, in accordance with the Galilean velocity transformation equation. If we assume the second alternative, we must abandon the notions of absolute time and absolute length that form the basis for the Galilean space-time transformation equations. … 
    
    The stage was set for Einstein, who solved the problem in 1905 with his special theory of relativity.
\end{quote}
Along with other physicists as noted above\cite{norton2004}, Einstein first tried and failed to produce a constructive theory for time dilation and length contraction as needed to explain why everyone measures the same value for the speed of light $c$, regardless of their motion relative to the source. [He was dealing with electrodynamics, but these kinematic facts were germane to that effort\cite{norton2004}.] Concerning his decision to produce a principle explanation instead of a constructive explanation for time dilation and length contraction, Einstein writes\cite[pp. 51-52]{einstein1949}:
\begin{quote}
By and by I despaired of the possibility of discovering the true laws by means of constructive efforts based on known facts. The longer and the more despairingly I tried, the more I came to the conviction that only the discovery of a universal formal principle could lead us to assured results.
\end{quote}
That is\cite{mainwood2018}, ``there is no mention in relativity of exactly \textit{how} clocks slow, or \textit{why} meter sticks shrink'' (no ``constructive efforts''), nonetheless the principles of SR are so compelling that\cite{mainwood2018} ``physicists always seem so sure about the particular theory of Special Relativity, when so many others have been superseded in the meantime.'' 

Consequently, introductory physics textbooks introduce SR in purely principle fashion by noting that Einstein's relativity principle\cite[p. 1018]{serway}, ``The laws of physics must be the same in all inertial reference frames'' or ``no preferred reference frame'' (NPRF) for short, is generalized from Galileo's relativity principle\cite[p. 1013]{serway}, ``The laws of mechanics must be the same in all inertial frames of reference.'' Serway \& Jewett\cite[p. 1018]{serway} and Knight\cite[p. 1149]{knight} then show that NPRF entails the light postulate of SR, i.e., that everyone measures the same speed of light \textit{c}, regardless of their motions relative to the source. If there was only one reference frame for a source in which the speed of light equalled the prediction from Maxwell's equations ($c = \frac{1}{\sqrt{\mu_o\epsilon_o}}$), then that would certainly constitute a preferred reference frame. The mysteries of time dilation and length contraction are then understood to follow from an ``empirically discovered'' fact (light postulate), which itself follows from NPRF, rather than from any ``hypothetically constructed'' fact, such as the luminiferous aether. 

\begin{figure}
\begin{center}
\includegraphics [height = 55mm]{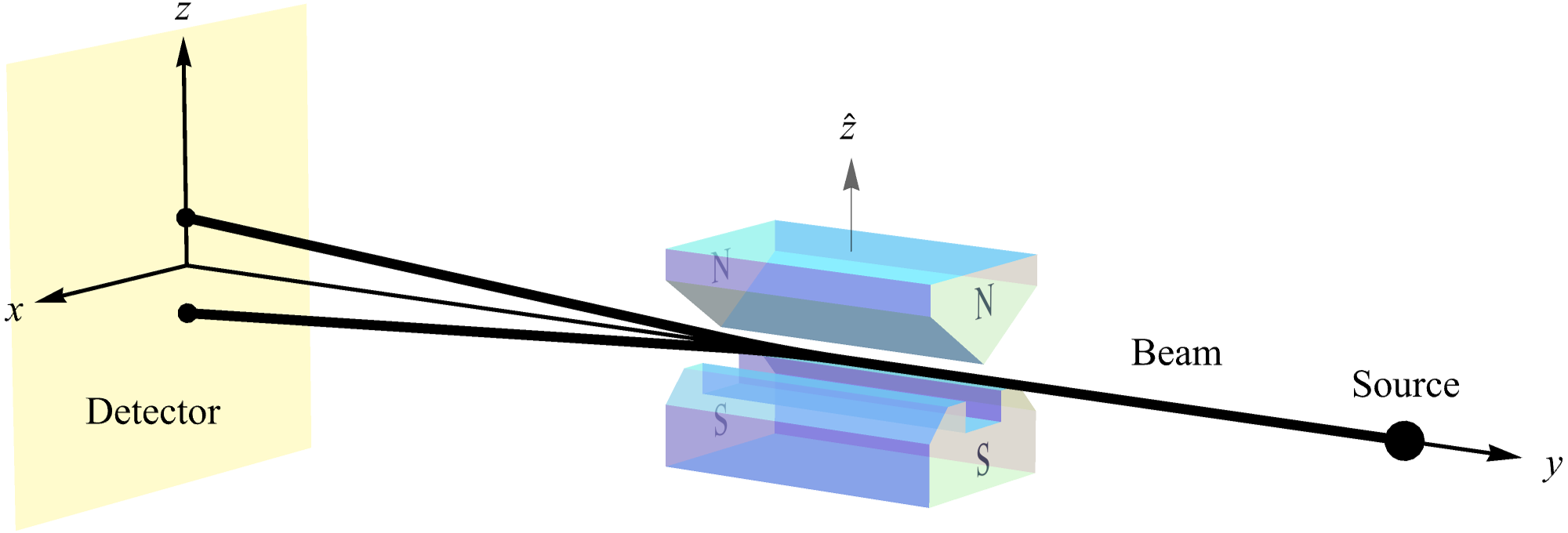}  \caption{A Stern-Gerlach (SG) spin measurement showing the two possible outcomes, up ($+\frac{\hbar}{2}$) and down ($-\frac{\hbar}{2}$) or $+1$ and $-1$, for short. As Weinberg points out, this constitutes a measurement of Planck's constant $h$.} \label{SGExp}
\end{center}
\end{figure}

With this typical introduction of SR in hand, the mystery of quantum entanglement can be introduced and resolved in principle fashion by further extending NPRF to include the measurement of another fundamental constant of nature, Planck's constant $h$. This can then be shown to result in `average-only' conservation of spin angular momentum characterizing the mystery of quantum entanglement per the uniquely quantum property of spin (see Appendix B for its generalization). At our institution, we cover the first two chapters on quantum physics in Serway \& Jewett\cite{serway} before introducing quantum entanglement so that the students have seen, among other things, Planck's radiation law, the Schr{\"o}dinger equation, and the photoelectric effect. Thus, they understand the importance of Planck's constant as a fundamental constant of Nature whose small-but-nonzero value is responsible for quantum physics as opposed to classical physics. [However, Planck obtained his radiation law and the value of $h$ first using classical physics alone\cite{PlanckRadiationLaw,PlanckRadiationYouTube}.] And, as Weinberg points out\cite{weinberg2017}, measuring an electron's spin via Stern-Gerlach (SG) magnets constitutes the measurement of ``a universal constant of nature, Planck's constant'' (Figure \ref{SGExp}). So if NPRF applies equally here, then everyone must measure the same value for Planck's constant \textit{h}, regardless of their SG magnet orientations relative to the source, which like the light postulate is an ``empirically discovered'' fact. By ``relative to the source,'' we might mean relative ``to the vertical in the plane perpendicular to the line of flight of the particles\cite[p. 943]{mermin1981},'' $\hat{z}$ in Figure \ref{SGExp} for example. 

Typically, only reference frames in relative motion at constant velocity are discussed when introducing the Lorentz transformations. The Lorentz transformations relating reference frames in relative motion at constant velocity are called \textit{Lorentz boosts}. However, spatial rotations are also part of the Lorentz transformations, indeed without them the Lorentz boosts do not form a group. [This does not mean rotational invariance implies Lorentz invariance, since spatial rotations are also part of the Galilean transformations.] For most physics experiments this is trivial, but for the SG measurement of Planck's constant $h$ we will see that this invariance proves significant. Thus, different SG magnet orientations relative to the source constitute different reference frames in QM just as different velocities relative to the source constitute different reference frames in SR. Indeed, an $[x,y,z]$ reference frame is naturally associated with the set of mutually complementary spin measurements $[J_x,J_y,J_z]$, so that two such reference frames are then naturally related by spatial rotation\cite{bruknerZeil2003} (Figure \ref{ComplBases}). Consequently, this principle resolution of the trilemma is as complete, satisfying, analytically rigorous, and accessible to first-year students as the standard introduction of SR. 

We begin in Section \ref{SecMerminDevice} by reviewing Mermin's famous introduction\cite{mermin1981} to the mystery of quantum entanglement via his ``Mermin device.'' Throughout the paper, we will work exclusively with spin-$\frac{1}{2}$ particles in a triplet state per the Mermin device, but the physics educator may of course convert to spin-$\frac{1}{2}$ particles in a singlet state\cite{MerminChallenge} or photons in the singlet state or triplet state\cite{dehlinger,janas2019,TsirelsonBound2019} if they prefer. In Section \ref{SecQM}, we then review the ``elementary quantum-mechanical reconciliation'' of cases (a) and (b) for the Mermin device (Mermin's wording\cite{mermin1981}). In Section \ref{SecNPRFandQM}, we show how that QM reconciliation, based on the mysterious `average-only' conservation, follows from NPRF applied to the SG measurement of Planck's constant $h$. We conclude in Section \ref{SecDiscussion} with further defense of this principle account.


\section{The Mermin Device: Introducing the Mystery} \label{SecMerminDevice}

\noindent As pointed out above, there are many ways to introduce the mystery of quantum entanglement to first-year or even general education (gen ed) students. Herein, we choose Mermin's famous 1981 introduction\cite{mermin1981} for the ``general reader,'' since it also contains an introduction to the Bell inequality and easily maps to the spin-$\frac{1}{2}$ triplet state. Feynman even complimented this paper in a letter to Mermin writing\cite[p. 366-7]{feynman}, ``One of the most beautiful papers in physics that I know of is yours in the American Journal of Physics.’’ As we will show, the mystery of quantum entanglement in this case, i.e., `average-only' conservation, is easy to resolve in principle fashion for the ``general reader'' via NPRF (Section \ref{SecNPRFandQM}).  

\begin{figure}[h!]
\begin{center}
\includegraphics [height = 50mm]{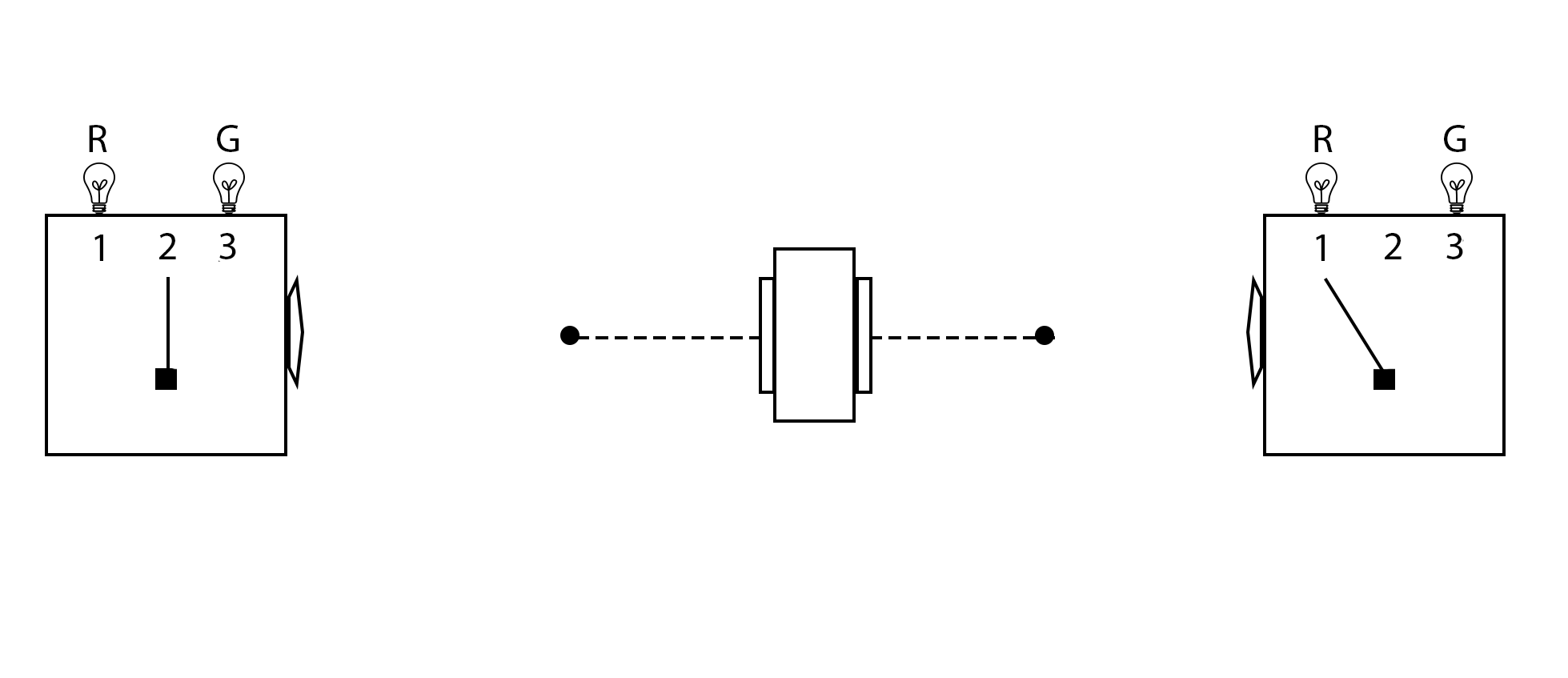}  \caption{\textbf{The Mermin Device}. Alice has her measuring device on the left set to 2 and Bob has his measuring device on the right set to 1. The particles have been emitted by the source in the middle and are in route to the measuring devices. Figure reproduced from Stuckey et al.\cite{MerminChallenge}} \label{mermin}
\end{center}
\end{figure}

\begin{figure}[h!]
\begin{center}
\includegraphics [height = 40mm]{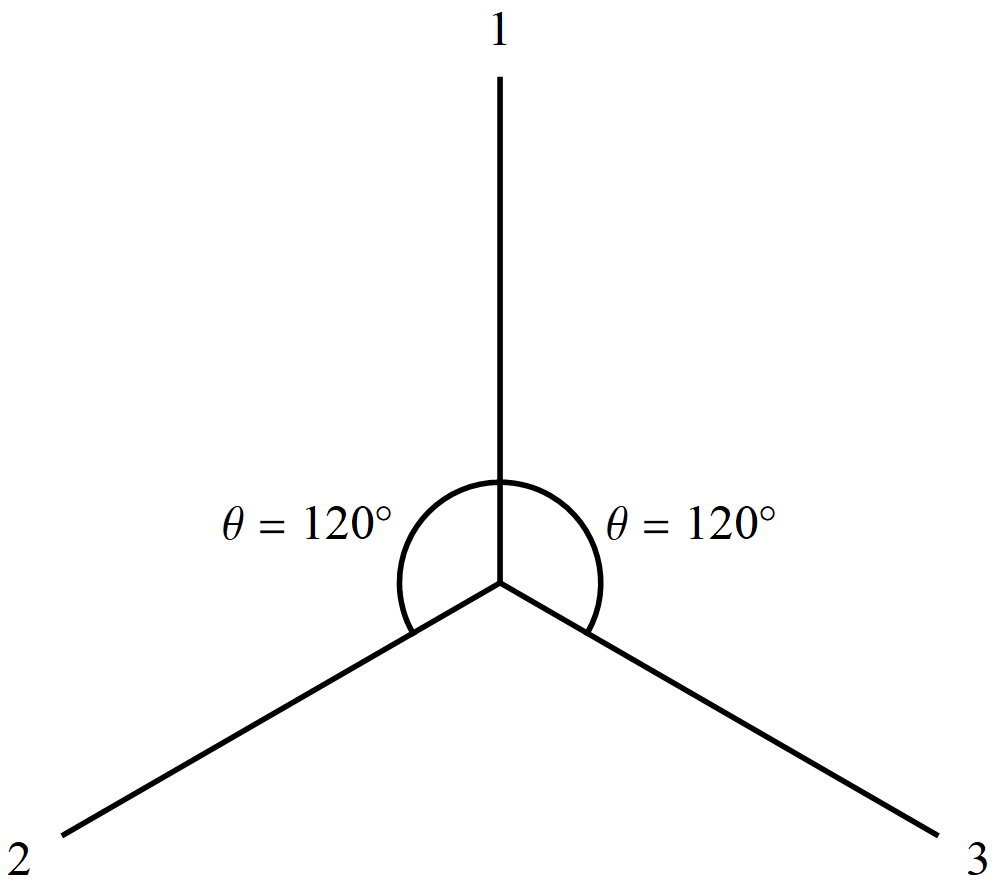} 
\caption{Three possible orientations of Alice and Bob's SG magnets for the Mermin device. Figure reproduced from Stuckey et al.\cite{MerminChallenge}} \label{SGorientations}
\end{center}
\end{figure}

The Mermin device contains a source (middle box in Figure \ref{mermin}) that emits a pair of spin-entangled particles towards two detectors (boxes on the left and right in Figure \ref{mermin}) in each trial of the experiment. The settings (1, 2, or 3) on the left and right detectors are controlled randomly by Alice and Bob. Each measurement at each detector produces either a result of R or G. These are the two facts that produce the mystery (Table \ref{tb:1}):
\begin{enumerate}
    \item[Fact 1.] When Alice and Bob's settings happen to be the same in a given trial (``case (a)''), their outcomes are always the same, $\frac{1}{2}$ of the time RR (Alice's outcome is R and Bob's outcome is R) and $\frac{1}{2}$ of the time GG (Alice's outcome is G and Bob's outcome is G).
    \item[Fact 2.] When Alice and Bob's settings happen to be different in a given trial (``case (b)''), the outcomes are the same $\frac{1}{4}$ of the time, $\frac{1}{8}$ RR and $\frac{1}{8}$ GG. 
\end{enumerate}
The two possible outcomes R and G represent the two possible spin measurement outcomes ``up'' and ``down,'' respectively (Figure \ref{SGExp}), and the three possible settings represent three different orientations of the SG magnets (Figures \ref{EPRBmeasure} \& \ref{SGorientations}). Mermin writes\cite[p. 942]{mermin1981}: 
\begin{quote}
Why do the detectors always flash the same colors when the switches are in the same positions? Since the two detectors are unconnected there is no way for one to ``know'' that the switch on the other is set in the same position as its own.
\end{quote}
Mermin introduces ``instruction sets'' to account for the outcomes when the detectors have the same settings. Concerning the use of instruction sets to account for Fact 1 he writes\cite[p. 942]{mermin1981}, ``It cannot be proved that there is no other way, but I challenge the reader to suggest any.'' Mermin has two constraints on those wishing to explain Facts 1 and 2. First, the particles cannot `know' what settings they will encounter until they arrive at the detectors (or, more generally, no violation of statistical independence). Second, they cannot communicate their settings and outcomes with each other in faster-than-light fashion (no violation of locality). At our institution, we have a class activity on the Mermin device, so we impose these constraints on that student activity (Appendix A). It doesn't take the students long to discover the need for instructions sets to account for Fact 1. This shows the students that Mermin's instruction sets can be understood to represent some constructive account of Fact 1 per his constraints. 

\begin{table}
\begin{center}
\begin{tabular}{ccc}
\textbf{Case (a) Same Settings} && \textbf{Case (b) Different Settings}\\
    \begin{tabular}{cc|cc}
    \multicolumn{2}{c}{}&\multicolumn{2}{c}{Alice} \\
    && R & G \\
    \cline{2-4}
    \multirow{2}*{Bob} & R & 1/2 & 0 \\
    & G &  0 & 1/2 
    \end{tabular}
& \hspace*{0.5in} &
    \begin{tabular}{cc|cc}
    \multicolumn{2}{c}{}&\multicolumn{2}{c}{Alice} \\
    && R & G \\
    \cline{2-4}
    \multirow{2}*{Bob} & R & 1/8 & 3/8 \\
    & G &  3/8 & 1/8 \\
    \end{tabular} \\
\end{tabular}
\caption{\textbf{Summary of outcome probabilities for the Mermin device.} Table reproduced from Stuckey et al.\cite{MerminChallenge}}
\label{tb:1}
\end{center}
\end{table}

\begin{table}
\begin{center}
\begin{tabular}{ccc}
\textbf{Case (a) Same Settings} && \textbf{Case (b) Different Settings}\\
\begin{tabular}{cc|cc}
    \multicolumn{2}{c}{}&\multicolumn{2}{c}{Alice} \\
    && R & G \\
    \cline{2-4}
    \multirow{2}*{Bob} & R & 1/2 & 0 \\
    & G &  0 & 1/2 
    \end{tabular}
& \hspace*{0.5in} &
    \begin{tabular}{cc|cc}
    \multicolumn{2}{c}{}&\multicolumn{2}{c}{Alice} \\
    && R & G \\
    \cline{2-4}
    \multirow{2}*{Bob} & R & 1/4 & 1/4 \\
    & G &  1/4 & 1/4 \\
    \end{tabular} \\
\end{tabular}
\caption{\textbf{Summary of outcome probabilities for instruction sets}. We are assuming the eight possible instruction sets are produced with equal frequency. Table reproduced from Stuckey et al.\cite{MerminChallenge}}
\label{tb:2}
\end{center}
\end{table}

Now consider all trials when Alice and Bob's particles have the instruction set GGR, for example. That means Alice and Bob's outcomes in setting 1 will both be G, in setting 2 they will both be G, and in setting 3 they will both be R. That is, the particles will produce a GG result when Alice and Bob both choose setting 1 (referred to as ``11''), a GG result when both choose setting 2 (referred to as ``22''), and an RR result when both choose setting 3 (referred to as ``33''). That is how instruction sets guarantee Fact 1. For different settings Alice and Bob will obtain the same outcomes when Alice chooses setting 1 and Bob chooses setting 2 (referred to as ``12''), which gives a GG outcome. And, they will obtain the same outcomes when Alice chooses setting 2 and Bob chooses setting 1 (referred to as ``21''), which also gives a GG outcome. That means we have the same outcomes for different settings in 2 of the 6 possible case (b) situations, i.e., in $\frac{1}{3}$ of case (b) trials for this instruction set. This $\frac{1}{3}$ ratio holds for any instruction set with two R(G) and one G(R). 

The only other possible instruction sets are RRR or GGG where Alice and Bob's outcomes will agree in $\frac{9}{9}$ of all trials. Thus, the ``Bell inequality'' for the Mermin device says that instruction sets must produce the same outcomes in more than $\frac{1}{3}$ of all case (b) trials\cite{alford2016}. Indeed, if all eight instruction sets are produced with equal frequency, the RR, GG, RG, and GR outcomes for any given pair of unlike settings (12, 13, 21, 23, 31, or 32) will be produced in equal numbers, so the probability of getting the same outcomes for different settings is $\frac{1}{2}$ (Table \ref{tb:2}). This fact is also discovered empirically by the students in the class activity. But, Fact 2 for QM says you only get the same outcomes in $\frac{1}{4}$ of all those trials, thereby violating the prediction per instruction sets, i.e., violating the Bell inequality. Thus, the mystery of quantum entanglement per the Mermin device is that the instruction sets (constructive account) needed for Fact 1 fail to yield the proper outcomes for Fact 2. And, a simple class activity based on the Mermin device makes the mystery of quantum entanglement tangible even for gen ed students (Appendix A).


\section{Elementary QM Reconciliation of Cases (\lowercase{a}) and (\lowercase{b})} \label{SecQM}

\noindent In this section, we review how the qubit Hilbert space structure of QM with its Pauli matrices reconciles Facts 1 and 2 concerning cases (a) and (b) for the Mermin device. At our institution, most of this material is reserved for those students who have contracted the course for Honors. However, all first-year students are shown Eqs. (\ref{QMavg}-\ref{QM-}), as necessary for Section \ref{SecNPRFandQM}, and the joint probabilities Eqs. (\ref{QM1jointLike}-\ref{QM2jointUnlike}) to convey the ``elementary quantum-mechanical reconciliation'' of cases (a) and (b). 

\clearpage

\begin{figure}
\begin{center}
\includegraphics [height = 55mm]{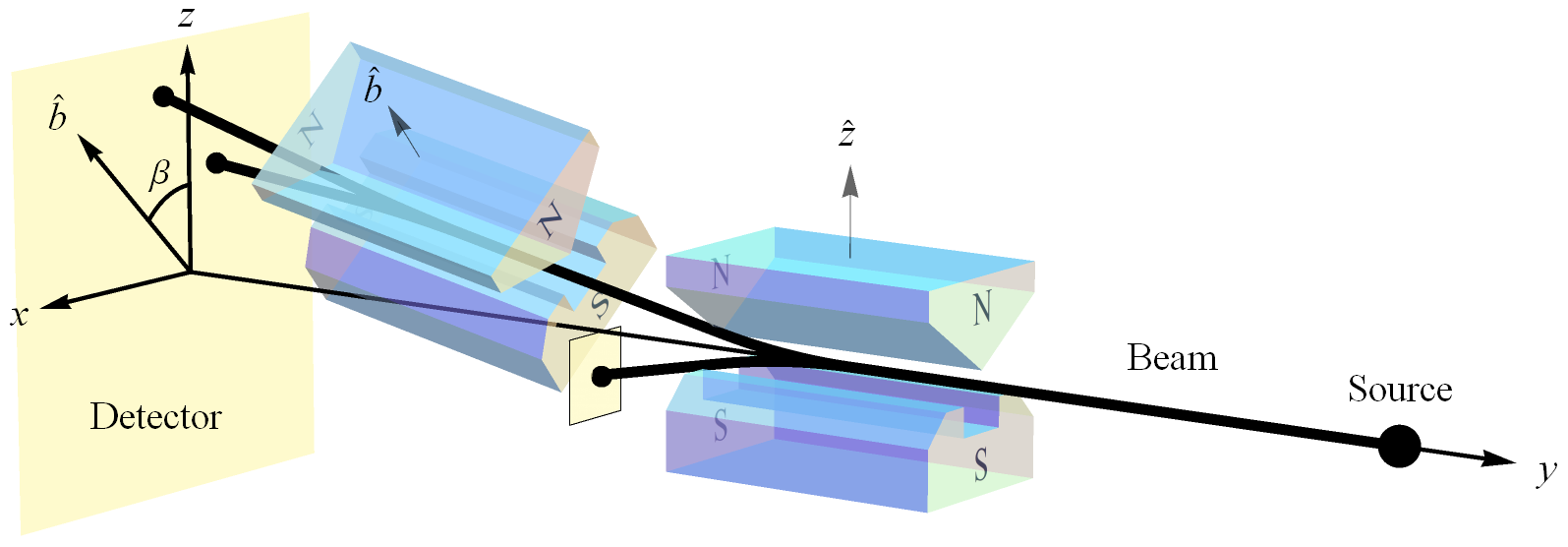}  \caption{In this set up, the first SG magnets (oriented at $\hat{z}$) are being used to produce an initial state $|\psi\rangle = |u\rangle$ for measurement by the second SG magnets (oriented at $\hat{b}$). Figure reproduced from Silberstein et al.\cite{silberstein2021}} \label{SGExp2}
\end{center}
\end{figure}

We will use the Dirac notation introduced by Ross\cite{ross2020}. In the eigenbasis of $\sigma_z$ the Pauli matrices are
$$
\sigma_x = \left ( \begin{array}{rr} 0 & \phantom{00}1 \\ 1 & \phantom{0}0 \end{array} \right ), \quad
\sigma_y = \left ( \begin{array}{rr} 0 & \phantom{0}-\textbf{i} \\ \textbf{i} & 0  \end{array} \right ), \quad \mbox{and} \quad
\sigma_z =\left ( \begin{array}{rr} 1 & 0 \\ 0 & \phantom{0}-1 \end{array} \right ).
$$
where $\textbf{i} = \sqrt{-1}$. The spin matrices $J_i = \frac{\hbar}{2}\sigma_i$ have the same eigenvalues (measurement outcomes) of $\pm \frac{\hbar}{2}$ in accord with Figure \ref{SGExp}. We will use the Pauli matrices with eigenvalues $\pm 1$ for short, and denote the corresponding eigenvectors (eigenstates) as $|u\rangle$ and $|d\rangle$ for spin up ($+1$) and spin down ($-1$), respectively. Using the Pauli matrices above with $|u\rangle = \left ( \begin{array}{rr} 1 \\ 0\end{array} \right )$ and $|d\rangle = \left ( \begin{array}{rr} 0 \\ 1\end{array} \right )$, we see that $\sigma_z|u\rangle = |u\rangle$, $\sigma_z|d\rangle = -|d\rangle$, $\sigma_x|u\rangle = |d\rangle$, $\sigma_x|d\rangle = |u\rangle$, $\sigma_y|u\rangle = \textbf{i}|d\rangle$, and $\sigma_y|d\rangle = -\textbf{i}|u\rangle$. If we flip the orientation of a vector from right pointing (ket) to left pointing (bra) or vice-versa, we transpose and take the complex conjugate. For example, if $|A\rangle = \textbf{i}\begin{pmatrix} 1\\0 \end{pmatrix} = \textbf{i}|u\rangle$, then $\langle A| = -\textbf{i}\begin{pmatrix} 1\;\; 0 \end{pmatrix} = -\textbf{i}\langle u|$. Thus, any spin matrix can be written as $(+1)|\tilde{u}\rangle\langle \tilde{u}| + (-1)|\tilde{d}\rangle\langle \tilde{d}|$ where $|\tilde{u}\rangle$ and $|\tilde{d}\rangle$ are its up and down eigenstates, respectively. A two-level quantum state $|\psi\rangle$ called a ``qubit'' is then given by $|\psi\rangle = c_1|u\rangle + c_2|d\rangle$, where $|c_1|^2 + |c_2|^2 = 1$. An arbitrary spin measurement $\sigma$ in the $\hat{b}$ direction is given by the Pauli matrices 
\begin{equation}
    \sigma = \hat{b}\cdot\vec{\sigma}=b_x\sigma_x + b_y\sigma_y + b_z\sigma_z \label{GenSigma}
\end{equation}
The average outcome for a measurement $\sigma$ on state $|\psi\rangle$ is given by
\begin{equation}
    \langle \sigma \rangle := \langle \psi | \sigma | \psi \rangle \label{AvgSigma}
\end{equation}
QM does not supply any means of predicting an exact outcome for any given trial, unless the probability happens to be one for that particular outcome in that particular configuration. As Mermin points out\cite[p. 10]{mermin2019}:
\begin{quote}
Quantum mechanics is, after all, the first physical theory in which probability is explicitly not a way of dealing with ignorance of the precise values of existing quantities.
\end{quote} 
As we will see in Section \ref{SecNPRFandQM}, this unavoidably probabilistic nature of QM gives rise to a mysterious `average-only' conservation for the Bell states precisely as needed to reconcile cases (a) and (b). Thus, the QM reconciliation of cases (a) and (b) is itself mysterious. Continuing, suppose $|\psi\rangle = |u\rangle$ (prepared by the first SG magnets in Figure \ref{SGExp2}) and $\sigma = \sin{(\beta)}\sigma_x + \cos{(\beta)}\sigma_z$ (per the second SG magnets in Figure \ref{SGExp2}). Using the Dirac formalism above, it is then easy to compute 
\begin{equation}
\langle \sigma \rangle = \cos{(\beta)} \label{QMavg}
\end{equation}
The probability of obtaining a $+1$ or $-1$ result for the measurement $\sigma$ is given by
\begin{equation}
P(+1 \mid \beta) = |\langle \psi | \tilde{u}\rangle|^2 = \mbox{cos}^2 \left(\frac{\beta}{2} \right) \label{QM+}
\end{equation}
and 
\begin{equation}
P(-1 \mid \beta) = |\langle \psi | \tilde{d}\rangle|^2 = \mbox{sin}^2 \left(\frac{\beta}{2} \right) \label{QM-}
\end{equation}
where $|\tilde{u}\rangle$ and $|\tilde{d}\rangle$ are the eigenvectors of $\sigma$. 

With that review of the relevant qubit Hilbert space structure, we now obtain the correlation functions for the Bell states, which will represent a spin-entangled pair of particles for us. The correlation function is how we will connect NPRF to the type of conservation represented by the Bell states. As we will see in Section \ref{SecNPRFandQM}, the type of conservation at work here is nothing like that in classical physics. That's because when Alice and Bob are making their measurements in different reference frames (at different SG magnet angles), spin angular momentum can only be conserved \textit{on average}, not on a trial-by-trial basis. Thus, this `average-only' conservation does not \textit{resolve} the mystery of quantum entanglement, it is simply another \textit{articulation of} the mystery. In principle, the creation of an entangled state due to conservation of spin angular momentum is not difficult to imagine, e.g., the dissociation of a spin-zero diatomic molecule\citep{bohm}. In reality, creating a Bell state in a controlled experimental situation is nontrivial\citep{hensen}, e.g., see Dehlinger \& Mitchell's detailed explanation for how they created a triplet state with photons\citep{dehlinger}.

\clearpage

As shown by Ross\cite{ross2020}, two-particle states are created from single-particle states using the tensor product $\otimes$, so that $\left(\sigma_x\otimes\sigma_z\right)\left(|u\rangle\otimes|d\rangle\right) = -|d\rangle\otimes|d\rangle$ and $\left(\sigma_x\otimes\sigma_y\right)\left(|u\rangle\otimes|d\rangle\right) = -\textbf{i}|d\rangle\otimes|u\rangle$, for example. In this notation, the Bell states are
\begin{equation}
\begin{aligned}
&|\psi_-\rangle = \frac{|u\rangle\otimes|d\rangle \,- |d\rangle\otimes|u\rangle}{\sqrt{2}}\\
&|\psi_+\rangle = \frac{|u\rangle\otimes|d\rangle + |d\rangle\otimes|u\rangle}{\sqrt{2}}\\
&|\phi_-\rangle = \frac{|u\rangle\otimes|u\rangle \,- |d\rangle\otimes|d\rangle}{\sqrt{2}}\\
&|\phi_+\rangle = \frac{|u\rangle\otimes|u\rangle + |d\rangle\otimes|d\rangle}{\sqrt{2}}\\ \label{BellStates}
\end{aligned}
\end{equation}
in the eigenbasis of $\sigma_z$. The first state $|\psi_-\rangle$ is called the \textit{singlet state} and it represents a total conserved spin angular momentum of zero ($S = 0$) for the two particles involved, i.e., Alice and Bob always obtain opposite outcomes ($ud$ or $du$) when measuring at the same angle. The other three states are called the \textit{triplet states} and they each represent a total conserved spin angular momentum of one ($S = 1$, in units of $\hbar = 1$), i.e., Alice and Bob always obtain the same outcomes ($uu$ or $dd$) when measuring at the same angle in the symmetry plane (see below). [To see this for $|\psi_+\rangle$, you have to transform the state to either the $\sigma_x$ or $\sigma_y$ eigenbasis where it has the same form as $|\phi_-\rangle$ or $|\phi_+\rangle$, respectively\cite{MerminChallenge}.] In all four cases, the entanglement represents the conservation of spin angular momentum for the process creating the state. 

\begin{figure}
\begin{center}
\includegraphics [height = 50mm]{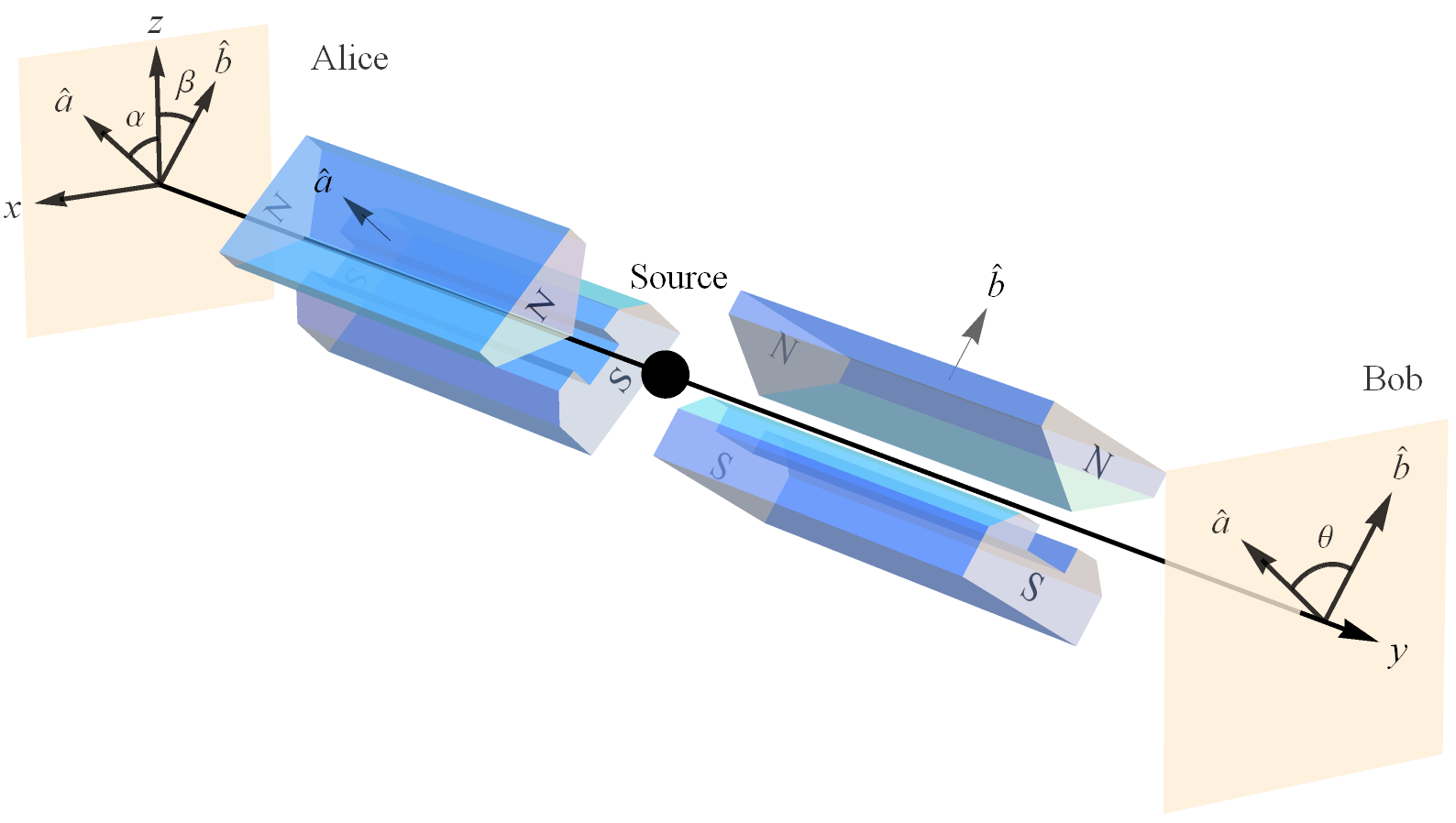}  \caption{Alice and Bob making spin measurements on a pair of spin-entangled particles in the Bell state $|\phi_+\rangle$ with their SG magnets and detectors. Figure reproduced from Silberstein et al.\cite{silberstein2021}} \label{EPRBmeasure}
\end{center}
\end{figure}

Suppose that Alice makes her spin measurement $\sigma_1$ in the $\hat{a}$ direction and Bob makes his spin measurement $\sigma_2$ in the $\hat{b}$ direction (Figure \ref{EPRBmeasure}), and let $\theta$ be the angle between $\hat{a}$ and $\hat{b}$.  Then
\begin{equation}
\begin{aligned}
    &\sigma_1 = \hat{a}\cdot\vec{\sigma}=a_x\sigma_x + a_y\sigma_y + a_z\sigma_z \\
    &\sigma_2 = \hat{b}\cdot\vec{\sigma}=b_x\sigma_x + b_y\sigma_y + b_z\sigma_z \\ \label{sigmas}
\end{aligned}
\end{equation}

\clearpage

\noindent Using the formalism explicated above, we can easily compute the correlation functions, i.e., the average of Alice and Bob's outcomes multiplied together for each Bell state
\begin{equation}
\begin{aligned}
&\langle\psi_-|\left(\sigma_1\otimes\sigma_2\right)|\psi_-\rangle = &-a_xb_x - a_yb_y - a_zb_z\\
&\langle\psi_+|\left(\sigma_1\otimes\sigma_2\right)|\psi_+\rangle = &a_xb_x + a_yb_y - a_zb_z\\
&\langle\phi_-|\left(\sigma_1\otimes\sigma_2\right)|\phi_-\rangle = &-a_xb_x + a_yb_y + a_zb_z\\
&\langle\phi_+|\left(\sigma_1\otimes\sigma_2\right)|\phi_+\rangle = &a_xb_x - a_yb_y + a_zb_z\\ \label{gencorrelations}
\end{aligned}
\end{equation}
The joint probabilities for Alice and Bob's measurements of a triplet state $\langle \psi_T|$ in its symmetry plane (Eq. (\ref{BellStates}) and Figure \ref{EPRBmeasure}) are given by
\begin{equation}
P(+1,+1 \mid \theta) = |\langle \psi_T | |u\rangle \otimes |\tilde{u} \rangle \rangle|^2 = \frac{1}{2} \mbox{cos}^2 \left(\frac{\theta}{2} \right)  \label{QM1jointLike}
\end{equation}
\begin{equation}
P(-1,-1 \mid \theta) = |\langle \psi_T | |d\rangle \otimes |\tilde{d}\rangle \rangle|^2 = \frac{1}{2} \mbox{cos}^2 \left(\frac{\theta}{2} \right)  \label{QM2jointLike}
\end{equation}
\begin{equation}
P(+1,-1 \mid \theta) = |\langle \psi_T | |u\rangle \otimes |\tilde{d}\rangle \rangle|^2 = \frac{1}{2} \mbox{sin}^2 \left(\frac{\theta}{2} \right) \label{QM1jointUnlike}
\end{equation}
\begin{equation}
P(-1,+1 \mid \theta) = |\langle \psi_T | |d\rangle \otimes |\tilde{u}\rangle \rangle|^2 = \frac{1}{2} \mbox{sin}^2 \left(\frac{\theta}{2} \right) \label{QM2jointUnlike}
\end{equation}
where $|u\rangle$ and $|d\rangle$ are the eigenvectors of $\sigma_1$ for Alice and $|\tilde{u}\rangle$ and $|\tilde{d}\rangle$ are the eigenvectors of $\sigma_2$ for Bob. These reconcile cases (a) and (b) for the Mermin device. That is, $\theta = 0$ for case (a) of Fact 1 means $P(+1,+1 \mid \theta) = P(-1,-1 \mid \theta) = \frac{1}{2}$ and $P(+1,-1 \mid \theta) = P(-1,+1 \mid \theta) = 0$, while $\theta = 120^{\circ}$ for case (b) of Fact 2 means $P(+1,+1 \mid \theta) = P(-1,-1 \mid \theta) = \frac{1}{8}$ and $P(+1,-1 \mid \theta) = P(-1,+1 \mid \theta) = \frac{3}{8}$ per Figure \ref{SGorientations} and Table \ref{tb:1}.

What does all this mean? The correlation function for $|\psi_-\rangle$ is $-\hat{a}\cdot\hat{b} = -\cos{(\theta)}$. This is invariant under rotations in any spatial plane, and that makes sense since the spin singlet state represents the conservation of a total spin angular momentum of $S = 0$, which is directionless. In contrast, the spin triplet states only have the rotationally invariant form $\cos{(\theta)}$ in a particular spatial plane (the ``symmetry plane''), as can be seen in Eq. (\ref{gencorrelations}). Using Eqs. (\ref{QM1jointLike})--(\ref{QM2jointUnlike}) to compute the correlation function in the symmetry plane of a triplet state we have
\begin{equation}
\begin{aligned}
&P(+1,+1 \mid \theta)(+1)(+1) + P(-1,-1 \mid \theta)(-1)(-1) + \\
&P(+1,-1 \mid \theta)(+1)(-1) + P(-1,+1 \mid \theta)(-1)(+1) = \\
&\frac{1}{2} \mbox{cos}^2 \left(\frac{\theta}{2}\right)(+1)(+1) + \frac{1}{2} \mbox{cos}^2 \left(\frac{\theta}{2} \right)(-1)(-1) + \\
&\frac{1}{2} \mbox{sin}^2 \left(\frac{\theta}{2} \right)(+1)(-1) + \frac{1}{2} \mbox{sin}^2 \left(\frac{\theta}{2} \right)(-1)(+1) = \\
&\mbox{cos}^2 \left(\frac{\theta}{2} \right) - \mbox{sin}^2 \left(\frac{\theta}{2} \right) = \cos{\left( \theta \right)} \label{TripletCorr}
\end{aligned}
\end{equation}
which agrees with Eq. (\ref{gencorrelations}). Thus, the spin triplet states represent conservation of spin angular momentum $S = 1$ in each of the spatial planes $xz$ ($|\phi_+\rangle$), $yz$ ($|\phi_-\rangle$), and $xy$ ($|\psi_+\rangle$). Specifically, when the SG magnets are aligned (the measurements are being made in the same reference frame) anywhere in the respective symmetry plane the outcomes are always the same ($\frac{1}{2}$ $uu$ and $\frac{1}{2}$ $dd$). If you want to model a conserved $S = 1$ for some other plane, you simply create a superposition, i.e., expand in the spin triplet basis. 

\begin{figure}[h]
\begin{center}
\includegraphics [width=\textwidth]{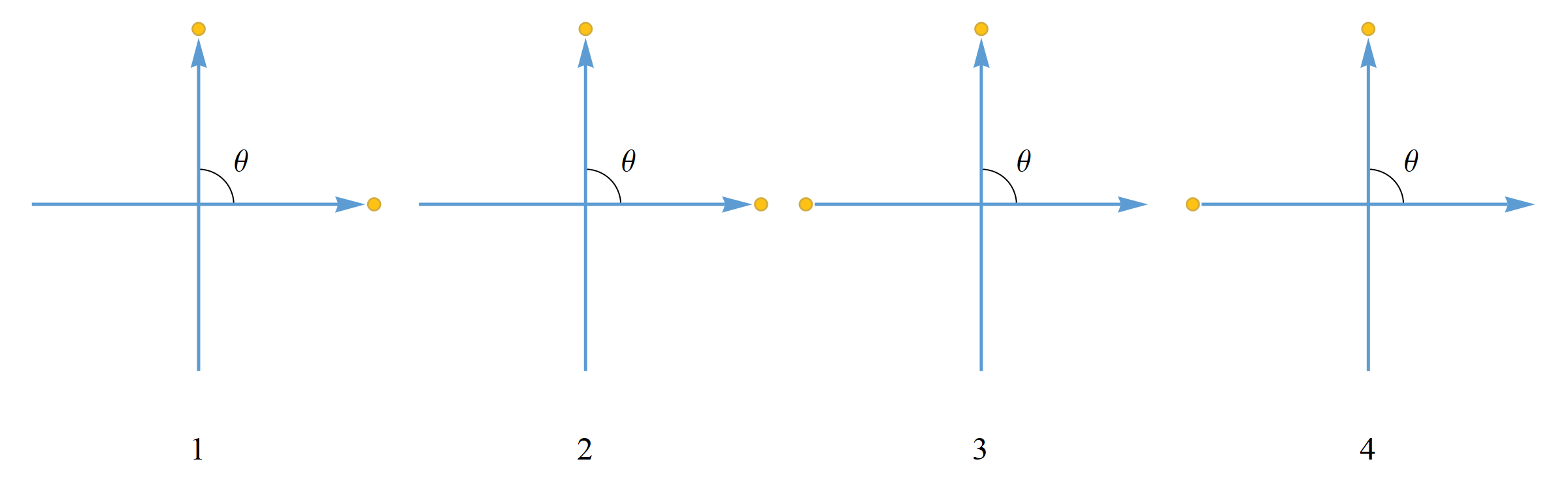}
\caption{An ensemble of 4 SG measurement trials of a spin triplet state showing Bob's(Alice's) outcomes corresponding to Alice's(Bob's) $+1$ outcome when $\theta = 90^\circ$. Spin angular momentum is not conserved in any given trial, because there are two different measurements being made, i.e., outcomes are in two different reference frames, but it is conserved \textit{on average} for all 4 trials (two up outcomes and two down outcomes average to $\cos{(90^\circ)}= 0$). At this value of $\theta$, Alice(Bob) says Bob's(Alice's) outcomes violate conservation of `intrinsic' angular momentum in maximal fashion on a trial-by-trial basis, since you can't obtain a result farther from zero than $\pm 1$. } \label{4Dpattern}
\end{center}
\end{figure}

While this conservation might seem prima facie to resolve the mystery of quantum entanglement, notice that what we said about the conservation of spin angular momentum deals only with SG measurements made at the same orientation (in the same reference frame). This corresponds to case (a) for the Mermin device and we showed that our instruction sets (representing some underlying constructive account) will easily produce the case (a) outcomes. The source of the mystery for the Mermin device was that our underlying constructive account failed to yield the case (b) outcomes. What we will next show is that the conservation depicted by this QM Hilbert space structure at \textit{different} SG measurement orientations can hold only \textit{on average}, not on a trial-by-trial basis. Indeed, the trial-by-trial outcomes for this `average-only' conservation can deviate substantially from the target value required for explicit conservation of spin angular momentum. For example, we might have $+1$ and $-1$ outcomes averaging to zero as required for the conservation of spin angular momentum (Figure \ref{4Dpattern}). In classical physics, our conservation laws hold on average because they hold explicitly for each and every trial of the experiment (within experimental limits). Thus, the conservation depicted by this QM reconciliation of cases (a) and (b) does not \textit{resolve} the mystery of quantum entanglement, it \textit{is} the mystery, i.e., it is what needs to be explained.


\section{The QM Reconciliation from NPRF} \label{SecNPRFandQM}

\noindent In this section, we detail our principle resolution of the mystery introduced in Section \ref{SecMerminDevice}. It is accessible to the first-year students in its entirety, while the conceptual parts and simple mathematics are accessible to the gen ed students. Of course, each physics educator will decide for themselves how to present the material for their students.

\begin{figure}
\begin{center}
\includegraphics [height = 65mm]{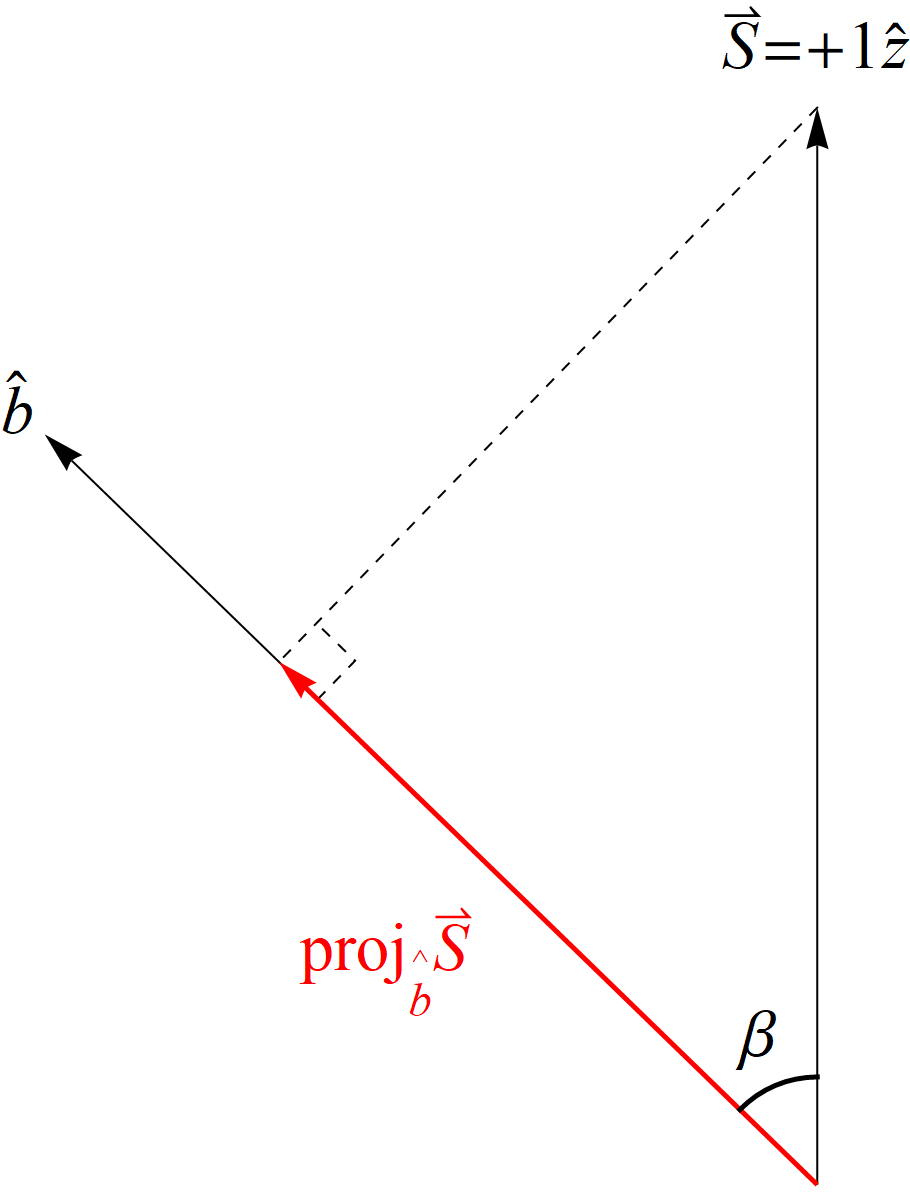} \caption{The `intrinsic' angular momentum of Bob's particle $\vec{S}$ projected along his measurement direction $\hat{b}$. This does \textit{not} happen with spin angular momentum due to NPRF. Figure reproduced from Silberstein et al.\cite{silberstein2021}} \label{Projection}
\end{center}
\end{figure}

\begin{figure}[h]
\begin{center}
\includegraphics [width=\textwidth]{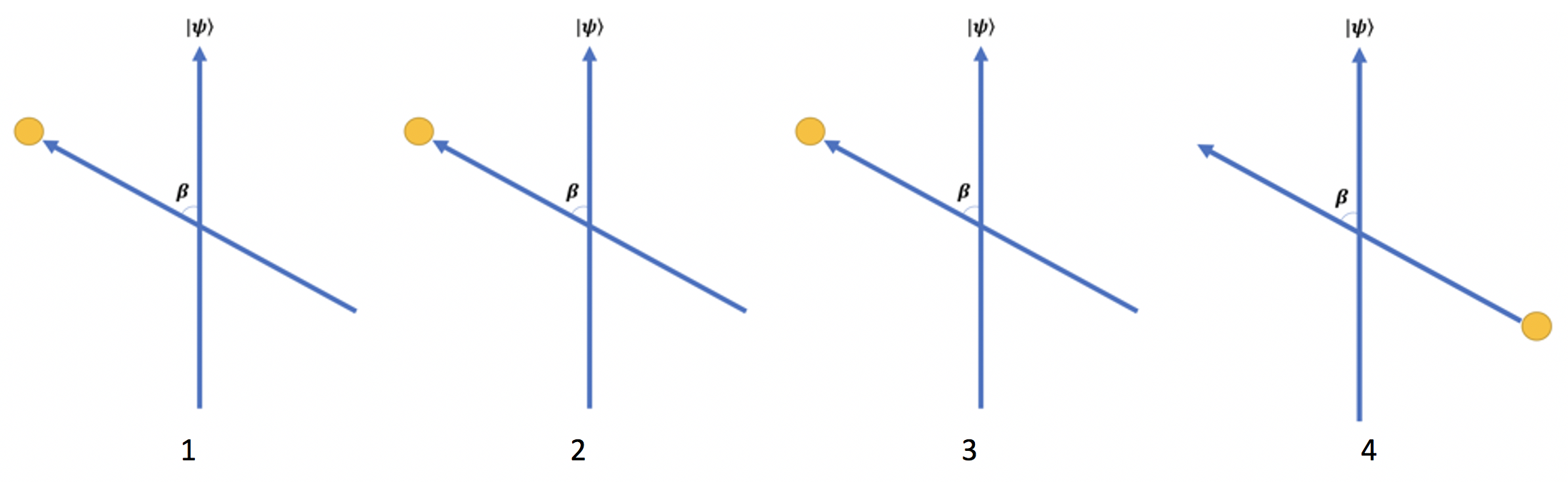} 
\caption{An ensemble of 4 SG measurement trials for $\beta = 60^{\circ}$ in Figure \ref{SGExp2}. The tilted blue arrow depicts the SG measurement orientation $\hat{b}$ and the vertical arrow represents our preparation state $|\psi\rangle = |u\rangle$. The yellow dots represent the two possible measurement outcomes for each trial, up (located at arrow tip) or down (located at bottom of arrow). The \textit{average} of the $\pm 1$ outcomes equals the projection of the initial spin angular momentum vector $\vec{S} = +1\hat{z}$ in the measurement direction $\hat{b}$, i.e.,  $\vec{S}\cdot\hat{b} = \cos{(60^\circ)}=\frac{1}{2}$.} \label{4DpatternBeta}
\end{center}
\end{figure}

As we considered above, suppose we produce a preparation of a quantum state oriented along the positive $z$ axis as in Figure \ref{SGExp2}, i.e., $|\psi\rangle = |u\rangle$, so that our `inherent/intrinsic' angular momentum is $\vec{S} = +1\hat{z}$ (in units of $\frac{\hbar}{2} = 1$). Now proceed to make a measurement with the SG magnets oriented at $\hat{b}$ making an angle $\beta$ with respect to $\hat{z}$ (Figure \ref{SGExp2}). According to the constructive account of classical physics\cite{knight,franklin2019} (Figure \ref{SGclassical}), we expect to measure $\vec{S}\cdot\hat{b} = \cos{(\beta)}$ (Figure \ref{Projection}), but we cannot measure anything other than $\pm 1$ due to NPRF (contra the prediction by classical physics). As a consequence, we can only recover $\cos{(\beta)}$ \textit{on average} (Figure \ref{4DpatternBeta}), i.e., NPRF dictates `average-only' projection
\begin{equation}
(+1) P(+1 \mid \beta) + (-1) P(-1 \mid \beta) = \cos (\beta) \label{AvgProjection}
\end{equation}
Thus, NPRF explains Eq. (\ref{QMavg}) of the qubit Hilbert space structure by providing a principle explanation of the uniquely quantum property of spin. Interestingly, it was years before the Stern-Gerlach experiment of 1922 was explained via spin, since the quantum prediction was based erroneously on the atom's orbital angular momentum\cite{friedrich2003}:
\begin{quote}
    However, the earliest attribution of the splitting to spin that we have found did not appear until 1927, when Ronald Fraser noted that the ground-state orbital angular momentum and associated magnetic moments of silver, hydrogen, and sodium are zero. Practically all current textbooks describe the Stern–Gerlach splitting as demonstrating electron spin, without pointing out that the intrepid experimenters had no idea it was spin that they had discovered. 
\end{quote}
Eq. (\ref{AvgProjection}) with our normalization condition $P(+1 \mid \beta) +  P(-1 \mid \beta) = 1$ then gives 
\begin{equation}
P(+1 \mid \beta) = \mbox{cos}^2 \left(\frac{\beta}{2} \right) \label{UPprobability}
\end{equation}
and 
\begin{equation}
P(-1 \mid \beta) = \mbox{sin}^2 \left(\frac{\beta}{2} \right) \label{DOWNprobability}
\end{equation}
Thus, NPRF also explains Eqs. (\ref{QM+}) \& (\ref{QM-}) of the qubit Hilbert space structure. In short, NPRF explains the unavoidably probabilistic nature of QM (see Appendix B for the generalization) and even provides the exact functional form of those probabilities. 

Let us emphasize here again that these mathematical facts follow from the ``empirically discovered'' fact that ``everyone measures the same value for $h$, regardless of their SG magnet orientation relative to the source,'' which itself follows from NPRF. Thus, NPRF is the fundamental principle responsible for the uniquely quantum property of spin (more about that in Section \ref{SecDiscussion} and Appendix B). We now follow Einstein's lead and provide a principle account of quantum entanglement by showing how `average-only' projection leads to `average-only' conservation, both being underwritten by NPRF. Accordingly, we see that the mysterious `average-only' conservation at the heart of the QM reconciliation is conservation that obtains because of NPRF. Conservation per NPRF then accounts for the correlation function of Eq. (\ref{TripletCorr}) and the joint probabilities reconciling cases (a) and (b) for the Mermin device. 

The correlation function is the average of the product of the outcomes $i$ and $j$ in each trial, $i \cdot j$, for settings $\alpha$ and $\beta$. That is
\begin{equation}
\langle \alpha,\beta \rangle = \sum (i \cdot j) \cdot P(i,j \mid \alpha,\beta)  \label{average}
\end{equation}
where $P(i,j \mid \alpha,\beta)$ are the quantum joint probabilities of observing $\pm 1$ for each of $i$ and $j$, given angle $\alpha$ for $\hat{a}$ and angle $\beta$ for $\hat{b}$. Again, we'll look specifically at a spin-$\frac{1}{2}$ triplet state per the Mermin device where $\langle \alpha,\beta \rangle = \mbox{cos} \left(\theta\right)$ in the symmetry plane (the spin singlet state is analogous\cite{MerminChallenge}). 

We have two sets of data, Alice's set and Bob's set. They were collected in $N$ pairs (data events) with Bob's(Alice's) SG magnets at $\theta$ relative to Alice's(Bob's). We want to compute the correlation function for these $N$ data events which is

\begin{equation}
\langle \alpha,\beta \rangle =\frac{(+1)_A(-1)_B + (+1)_A(+1)_B + (-1)_A(-1)_B + ...}{N}
\end{equation}
for the sample observations $(i=+1, j=-1), (i=+1,j=+1), (i=-1,j=-1),\ldots$. Next divide the numerator into two equal subsets per Alice's $+1$ results and Alice's $-1$ results
\begin{equation}
\langle \alpha,\beta \rangle =\frac{(+1)_A(\sum \mbox{BA+})+(-1)_A(\sum \mbox{BA-})}{N}
\end{equation}
where $\sum \mbox{BA+}$ is the sum of all of Bob's results (event labels) corresponding to Alice's $+1$ result (event label) and $\sum \mbox{BA-}$ is the sum of all of Bob's results (event labels) corresponding to Alice's $-1$ result (event label). Now, we rewrite that equation as
\begin{equation}
\langle \alpha,\beta \rangle = \frac{1}{2}(+1)_A\overline{BA+} + \frac{1}{2}(-1)_A\overline{BA-} \label{consCorrel}
\end{equation}
with the overline denoting average. Notice this correlation function is independent of the formalism of QM, all we have assumed is that Alice and Bob measure $+1$ or $-1$ with equal frequency at any setting in computing this correlation function (per NPRF). Thus, to understand the quantum correlation function for a spin-$\frac{1}{2}$ triplet state, we need to understand the origin of $\overline{BA+}$ and $\overline{BA-}$ for the spin-$\frac{1}{2}$ triplet state. 

\begin{table}
$$
    \begin{array}{cc|cc|c}
    \multicolumn{2}{c}{}& \multicolumn{2}{c}{\mbox{Bob}} \\
    && +1 & -1 & \mbox{Total}  \\
    \cline{2-5}
   \multirow{2}*{Alice} & +1 & P(+1,+1 \mid \theta) & P(+1,-1 \mid \theta) & 1/2 \\
    & -1 & P(-1,+1 \mid \theta) & P(-1,-1 \mid \theta) & 1/2 \\
     \cline{2-5}
    & \mbox{Total} & 1/2 & 1/2 & 1
    \end{array}
$$
\caption{{\bf Joint probabilities for Alice and Bob's outcome pairs for the entangled particle experiment in Figure \ref{EPRBmeasure} given an angle $\theta$.}  The table is symmetric due to NPRF.}
\label{tab:pmf}
\end{table}

\begin{figure}
\begin{center}
\includegraphics [height = 65mm]{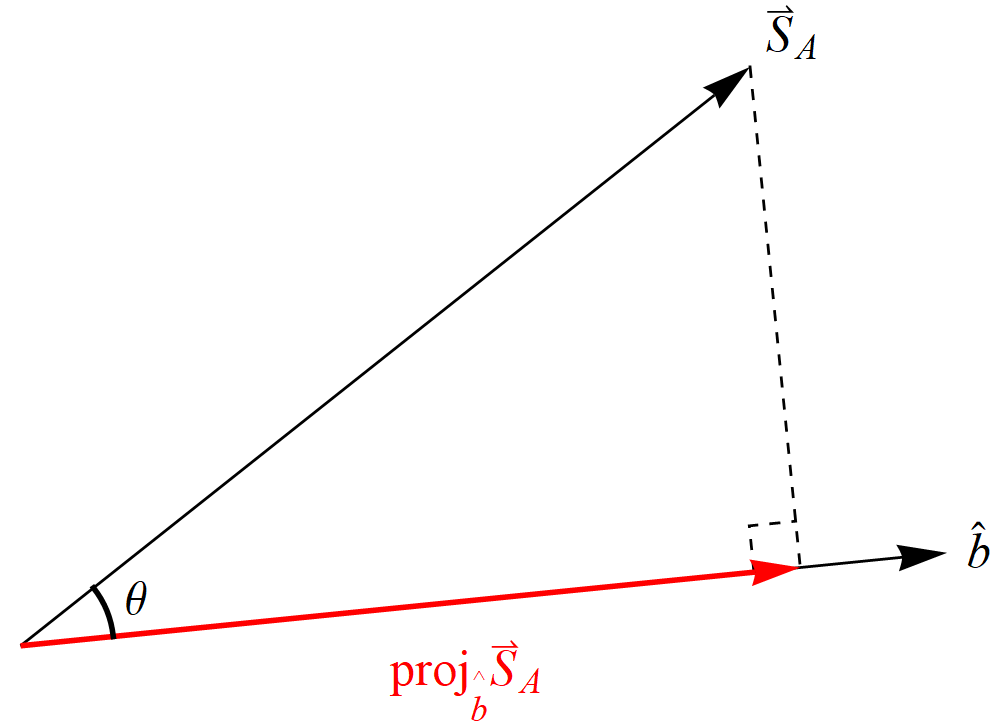} \caption{The `intrinsic' angular momentum of Bob's particle $\vec{S}_B = \vec{S}_A$ projected along his measurement direction $\hat{b}$. This does \textit{not} happen with spin angular momentum due to NPRF. Figure reproduced from Stuckey et al.\cite{MerminChallenge}} \label{TripletProjection}
\end{center}
\end{figure}

\begin{figure}
\begin{center}
\includegraphics [height = 24mm]{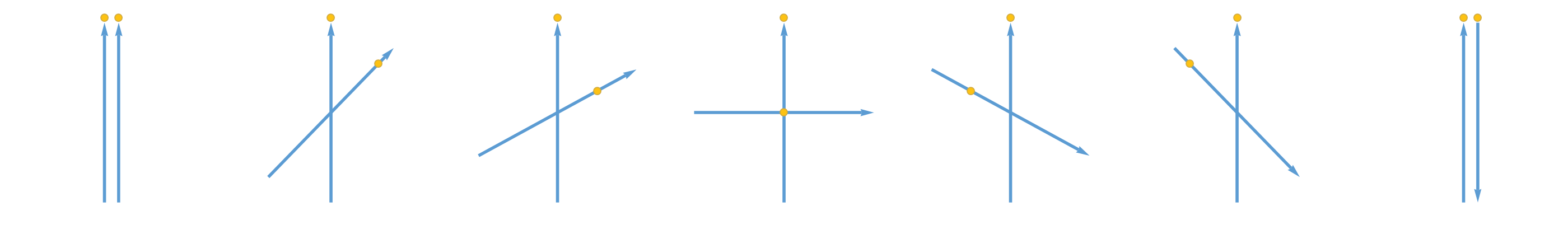}  \caption{\textbf{Average View for the Spin Triplet State}. Reading from left to right, as Bob rotates his SG magnets (rotating blue arrow) relative to Alice's SG magnets (blue arrow always vertically oriented) for her $+1$ outcome, the average value of his outcome varies from $+1$ (totally up, arrow tip) to $0$ to $-1$ (totally down, arrow bottom). This obtains per conservation of spin angular momentum on average in accord with NPRF. Bob can say exactly the same about Alice's outcomes as she rotates her SG magnets relative to his SG magnets for his $+1$ outcome. Figure reproduced from Silberstein et al.\cite{silberstein2021}} \label{AvgViewTriplet}
\end{center}
\end{figure}

As with the single-particle state, our constructive account per Figure \ref{SGclassical} would lead us to naively expect the projection of the `intrinsic' angular momentum vector of Alice's particle $\vec{S}_A = +1\hat{a}$ along $\hat{b}$ is $\vec{S}_A\cdot\hat{b} = +\cos({\theta})$ (Figure \ref{TripletProjection}) where $\theta$ is the angle between the unit vectors $\hat{a}$ and $\hat{b}$ (Figure \ref{EPRBmeasure}). Again, that's because the prediction from classical physics is that all values between $+1 \left(\frac{\hbar}{2}\right)$ and $-1 \left(\frac{\hbar}{2}\right)$ are possible outcomes for an `intrinsic' angular momentum measurement (Figure \ref{SGclassical}), i.e., $h \longrightarrow 0$ takes quantum physics to classical physics (Appendix B). From Alice's perspective, had Bob measured at the same angle, i.e., oriented his SG magnets in the same direction, he would have found the `intrinsic' angular momentum vector of his particle was $\vec{S}_B = \vec{S}_A = +1\hat{a}$ per conservation of angular momentum. Since he did not measure the `intrinsic' angular momentum of his particle at the same angle, he should have obtained a fraction of the length of $\vec{S}_B$, i.e., $\vec{S}_B\cdot\hat{b} = +1\hat{a}\cdot\hat{b} = \cos{(\theta)}$ (Figure \ref{TripletProjection}). But according to NPRF, Bob only ever obtains $+1$ or $-1$ just like Alice, so he cannot measure the required fractional outcome to explicitly conserve `intrinsic' angular momentum per Alice. Therefore, as with the single-particle case, Bob's outcomes can only satisfy the required projection \textit{on average} per NPRF (Figures \ref{4DpatternBeta}, \ref{4Dpattern}, \& \ref{AvgViewTriplet}), which means
\begin{equation}
\overline{BA+} = \mbox{cos} \left(\theta\right) \label{AGC1}
\end{equation}
Given this constraint per NPRF, as with the single-particle case, we can now use NPRF to find the joint probabilities for Alice and Bob's outcome pairs. Looking at Table \ref{tab:pmf}, the rows and columns all sum to $1/2$ because both Alice and Bob must observe $+1$ half of the time and $-1$ half of the time per NPRF, which also requires that the table is symmetric so that $P(-1,+1\mid \theta) = P(+1,-1\mid \theta)$.  The conditional distribution for Bob's outcome given that Alice observes a $+1$ is the top row in Table \ref{tab:pmf} divided by the row sum (1/2), so the average of Bob's outcomes given that Alice observes a $+1$ is
\begin{equation}
\overline{BA+} = 2P(+1,+1\mid \theta)(+1) + 2P(+1,-1\mid \theta)(-1) = \cos (\theta) \label{BA+}
\end{equation}
using conservation per NPRF. Together with the NPRF constraints on the rows/columns 
\begin{align*}
P(+1,+1\mid\theta) + P(+1,-1\mid \theta) & = \frac 12 \\
P(+1,-1\mid\theta) + P(-1,-1\mid \theta) & = \frac 12
\end{align*}
we can uniquely solve for the joint probabilities
\begin{equation}
P(+1,+1 \mid \theta) = P(-1,-1 \mid \theta) = \frac{1}{2} \mbox{cos}^2 \left(\frac{\theta}{2} \right) \label{QMjointLike}
\end{equation}
and 
\begin{equation}
P(+1,-1 \mid \theta) = P(-1,+1 \mid \theta) = \frac{1}{2} \mbox{sin}^2 \left(\frac{\theta}{2} \right) \label{QMjointUnlike}
\end{equation}
These agree with Eqs. (\ref{QM1jointLike})--(\ref{QM2jointUnlike}) for the QM reconciliation of cases (a) and (b) shown in Section \ref{SecQM}. Now we can use these to compute
\begin{equation}
\overline{BA-} = 2P(-1,+1\mid \theta)(+1) + 2P(-1,-1\mid \theta)(-1) = -\cos (\theta) \label{BA-}
\end{equation}
Using Eqs. (\ref{BA+}) \& (\ref{BA-}) in Eq. (\ref{consCorrel}) we obtain
\begin{equation}
\langle \alpha,\beta \rangle = \frac{1}{2}(+1)_A(\mbox{cos} \left(\theta\right)) + \frac{1}{2}(-1)_A(-\mbox{cos} \left(\theta\right)) = \mbox{cos} \left(\theta\right) 
\end{equation}
which is precisely the correlation function for a spin triplet state in its symmetry plane found in Section \ref{SecQM} using the QM Hilbert space structure, Eq. (\ref{TripletCorr}). 

There are two important points to be made here. First, NPRF is being used to justify an ``empirically discovered'' fact, i.e., Alice and Bob both always measure $\pm 1$ (Appendix B). Second, this ``empirically discovered'' fact has the mathematical consequence of `average-only' conservation (Eq. (\ref{BA+})) yielding the joint probabilities Eqs. (\ref{QMjointLike}) \& (\ref{QMjointUnlike}) responsible for the QM reconciliation of cases (a) and (b). In other words, to paraphrase Einstein, ``we have an empirically discovered principle that gives rise to mathematically formulated criteria which the separate processes or the theoretical representations of them have to satisfy.'' That is why this principle account of quantum entanglement provides ``logical perfection and security of the foundations'' exactly as in SR. And, as can be seen using the axiomatic reconstructions of QM via information-theoretic principles, it is also quite general (Appendix B). Thus, we see how quantum entanglement follows from ``NPRF + $h$'' in precisely the same manner that time dilation and length contraction follow from ``NPRF + $c$.'' And, just like in SR, Bob could divide the data according to his $\pm 1$ results (per his reference frame) and claim that it is Alice who must average her results (obtained in her reference frame) to conserve spin angular momentum (Table \ref{tab:SRvsQM}).

\begin{table}[]
    \centering
    \begin{tabular}{|c|c|}
    \hline
        {\bf Special Relativity} & {\bf Quantum Mechanics} \\
        \hline
        Empirical Fact: Alice and Bob both  &  Empirical Fact: Alice and Bob both   \\
     measure $c$, regardless of their   &  measure $\pm 1 \left (  \frac {\hbar}2 \right )$, regardless of their SG  \\
       
        motion relative to the source &  orientation relative to the source \\
        \hline
        Consequence: Alice(Bob) says Bob(Alice) must   &  Consequence: Alice(Bob) says Bob(Alice) \\ 
        correct his(her) time and length measurements & must average his(her) $\pm 1$ results \\
        \hline
        
    \end{tabular}
    \caption{\textbf{Principle comparison of special relativity and quantum mechanics.} Because Alice and Bob both measure the same speed of light $c$, regardless of their motion relative to the source per NPRF, Alice(Bob) may claim that Bob's(Alice's) length and time measurements are erroneous and need to be corrected (length contraction and time dilation). Likewise, because Alice and Bob both measure the same values for spin angular momentum $\pm 1$ $\left(\frac{\hbar}{2}\right)$, regardless of their SG magnet orientation relative to the source per NPRF, Alice(Bob) may claim that Bob's(Alice's) individual $\pm 1$ values are erroneous and need to be corrected (averaged, Figures \ref{4Dpattern} \& \ref{AvgViewTriplet}).}
    \label{tab:SRvsQM}
\end{table}


\section{Discussion}\label{SecDiscussion}

\noindent The 2022 Nobel Prize in Physics was awarded\cite{nobel2022} ``for experiments with entangled photons, establishing the violation of Bell inequalities and pioneering quantum information science.'' And, as stated by Timmerman in ``Redesigning quantum information science education and training: The Chicago Quantum Exchange approach'' for session A29, ``Quantum Information Education,'' in the 2021 APS March Meeting\cite{APS}:
\begin{quote}
The rapidly evolving field of quantum information science has the power to transform cybersecurity, materials development, computing, and other areas of research and innovation.    
\end{quote}
Further, as pointed out by Barnes in ``An educational program to teach quantum information science to high-school students'' for that same session, germane to quantum information science are ``the principles of superposition, entanglement, and measurement in quantum mechanics.'' While quantum entanglement is an important concept for this exciting new field of physics, there is a trilemma associated with its mystery. In short, does the physics educator ignore the mystery, tell their students to ``Shut up and calculate!,'' or delve into murky quantum interpretations, such as Many Worlds or superluminal pilot waves? 

\begin{figure}
\begin{center}
\includegraphics [height = 75mm]{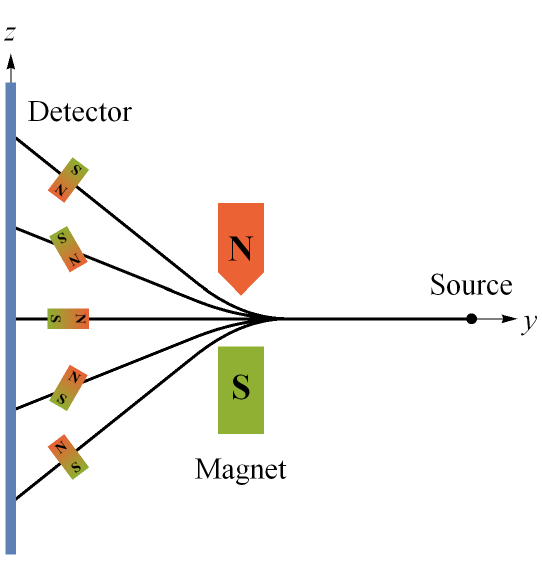}  \caption{\textbf{The classical constructive model of the Stern-Gerlach experiment.} If the atoms enter with random orientations of their `intrinsic' magnetic moments (due to their `intrinsic' angular momenta), the SG magnets should produce all possible deflections, not just the two that are observed. Figure reproduced from Stuckey et al.\cite{stuckey2022}} \label{SGclassical}
\end{center}
\end{figure}

Herein, we have presented a resolution of this pedagogical trilemma in exactly the same way that Einstein resolved the mysteries of time dilation and length contraction in 1905. Instead of finding a constructive account of time dilation and length contraction giving rise to the fact that everyone measures the same speed of light $c$, Einstein abandoned his ``constructive efforts'' and embraced a principle approach. That is, he reversed the explanatory order by using the relativity principle, i.e., ``no preferred reference frame'' (NPRF), to justify the light postulate giving rise to time dilation and length contraction. Following his lead, we abandoned ``constructive efforts'' to explain quantum entanglement and used NPRF to justify the fact that everyone measures the same value for Planck's constant $h$ responsible for the uniquely quantum property of spin angular momentum. The existence of spin angular momentum  then accounts for the `average-only' conservation of `intrinsic' angular momentum that characterizes the mystery of quantum entanglement per the Bell states. Thus, we have a principle resolution of Mermin's ``Quantum mysteries for anybody'' using spin-$\frac{1}{2}$ particles in the triplet state per the Mermin device. Accordingly, the Mermin device, Figures \ref{SGExp} -- \ref{AvgViewTriplet}, and Tables \ref{tb:1} -- \ref{tab:SRvsQM} allow the physics educator to conceptually introduce and resolve the mystery of quantum entanglement for first-year or gen ed students in precise analogy to the standard introduction of SR in introductory physics textbooks (Table \ref{tab:SRvsQM}). The mathematical details were supplied for completeness and might be shared with more advanced students at the discretion of the instructor.

Of course, this resolution of the mystery of quantum entanglement will only satisfy those who are likewise satisfied with the principle explanation of time dilation and length contraction per NPRF in SR. Even a conceptual introduction to physics can show how those mysteries follow from the light postulate\cite[pp. 442-445]{griffith} and that the light postulate follows from NPRF. But, as Lorentz complained\cite[p. 230]{lorentz}
\begin{quote}
Einstein simply postulates what we have deduced, with some difficulty and not altogether satisfactorily, from the fundamental equations of the electromagnetic field.
\end{quote}
And, Albert Michelson said\cite{michelson}
\begin{quote}
It must be admitted, these experiments are not sufficient to justify the hypothesis of an aether. But then, how can the negative result be explained?
\end{quote}
In other words, neither was convinced that the relativity principle was sufficient to explain time dilation and length contraction. Apparently for them, such a principle must be accounted for constructively, e.g., the luminiferous aether. However, the introductory physics textbooks offer no constructive counterpart to this principle account. In fact, they even (rightfully) dismiss as unreasonable the reigning constructive account of the late nineteenth century, viz., the luminiferous aether. 

But, someone who is not entirely satisfied with the principle resolution of the mysteries of time dilation and length contraction does not need to subscribe to the existence of a luminiferous aether. They might simply imagine there does exist an undiscovered Lorentz-invariant dynamics responsible for time dilation and length contraction, which then yield the light postulate\cite{brown2003}. However, even if such a dynamics exists, being Lorentz invariant, it would still obtain due to NPRF. As pointed out by Brown\cite[p. 76]{brownpooley2006}:
\begin{quote}
    What has been shown is that rods and clocks must behave in quite particular ways in order for the two postulates to be true together. But this hardly amounts to an explanation of such behaviour. Rather things go the other way around. It is because rods and clocks behave as they do, in a way that is consistent with the relativity principle, that light is measured to have the same speed in each inertial frame.
\end{quote} 
So, whether or not you are completely satisfied with the principle resolution of the mysteries of time dilation and length contraction presented in the introductory physics textbooks, you will still understand those mysteries to result from NPRF. In either case, everyone agrees that Einstein's principle account served to advance physics and we still don't have a (consensus) constructive counterpart. Even Lorentz seemed to acknowledge the value of this principle explanation when he wrote\cite[p. 230]{lorentz}
\begin{quote}
By doing so, [Einstein] may certainly take credit for making us see in the negative result of experiments like those of Michelson, Rayleigh, and Brace, not a fortuitous compensation of opposing effects but the manifestation of a general and fundamental principle.
\end{quote}  
Now let's look at the analogous situation concerning our principle resolution of the mystery of quantum entanglement.

Our counterpart to the empirically discovered light postulate is the empirically discovered fact that ``everyone measures the same value for Planck's constant $h$, regardless of the orientation of their SG magnets relative to the source.'' This is just another way to describe the uniquely quantum property of spin (Figure \ref{SGExp}), which the introductory physics textbooks point out also has no constructive account (Figure \ref{SGclassical}). And as we showed, this \textit{Planck postulate} underwrites the QM Hilbert space structure reconciling cases (a) and (b) for the Mermin device. So, have we simply postulated the existence of a mysterious property (spin angular momentum) to resolve the mystery of quantum entanglement? 

If we had stopped with our Planck postulate, the answer would be ``yes.'' But, we didn't stop with the Planck postulate any more than the standard introductory physics textbook presentation of SR stops with the light postulate. Rather, both empirically discovered facts are understood to obtain due to the relativity principle, NPRF. And, whatever dynamics is responsible for the $\pm 1 \left(\frac{\hbar}{2}\right)$ outcomes at some SG magnet orientation, since this SG measurement thereby constitutes a measurement of a fundamental constant of Nature, Planck's constant $h$, the Lorentz invariance of those dynamics, which includes spatial rotations, entails that those dynamics continue to produce $\pm 1$ outcomes as we rotate our SG magnets. Again, as with the light postulate, the quantum property of spin can be understood to result from NPRF whether you consider the principle account or some yet-to-be-discovered constructive account per Lorentz-invariant dynamics. Let us expand on that point.

The fact that one obtains $\pm 1$ outcomes at some SG magnet orientation is not mysterious per se, it can be accounted for by the classical constructive model in Figure \ref{SGclassical}. The constructive account of the $\pm 1$ outcomes would be one of particles with `intrinsic' angular momenta and therefore `intrinsic' magnetic moments\cite{knight} orientated in two opposite directions in space, parallel or anti-parallel to the magnetic field. Given this constructive account of the $\pm 1$ outcomes at this particular SG magnet orientation, we would then expect that the varying orientation of the SG magnetic field with respect to the magnetic moments, created as we rotate our SG magnets, would cause the degree of deflection to vary. Indeed, this is precisely the constructive account that led some physicists to expect all possible deflections for the particles as they passed through the SG magnets, having assumed that these particles would be entering the SG magnetic field with random orientations of their `intrinsic' magnetic moments\cite{franklin2019} (Figure \ref{SGclassical}). But according to this constructive account, if the $\pm 1$ outcomes constitute a measurement of $h$ in accord with the rest of quantum physics, then our rotated orientations would not be giving us the value for $h$ required by quantum physics otherwise. Indeed, a rotation of $90^\circ$ would yield absolutely no deflection at all (akin to measuring the speed of a light wave as zero when moving through the aether at speed $c$). That would mean our original SG magnet orientation would constitute a preferred frame in violation of the relativity principle, NPRF. Essentially, as Michelson and Morley rotated their interferometer the constructive model predicted they would see a change in the interference pattern\cite{michelson}, but instead they saw no change in the interference pattern in accord with NPRF. Likewise, as Stern and Gerlach rotated their magnets the constructive model predicted they would see a change in the deflection pattern, but instead they saw no change in the deflection pattern in accord with NPRF. 

So, we see that the quantum property of spin (our Planck postulate), which is yet today without a constructive account, can be understood to result in principle fashion from NPRF. And, as pointed out above, any constructive counterpart that may ultimately be produced would be the result of Lorentz-invariant dynamics. Therefore, either way, as with the mysteries of time dilation and length contraction, the mystery of quantum entanglement can be understood to result from the relativity principle, NPRF. 

It is true that we have not provided a constructive resolution of the mystery of quantum entanglement any more than Einstein did for the mysteries of time dilation and length contraction, but our principle resolution is certainly not without value, as even Lorentz acknowledged about SR. At very least, our principle resolution of the mystery of quantum entanglement adds a new option to the pedagogical arsenal of the physics educator interested in resolving the trilemma associated with that mystery. And, there is no reason to believe that a constructive counterpart is anywhere on the horizon. The famous Einstein, Podolsky, and Rosen paper\cite{EPRpaper} introducing the mystery of quantum entanglement was published in 1935, yet we still have no (consensus) constructive account of quantum entanglement. Therefore, as with SR, physics educators should consider the possibility that quantum entanglement will ultimately yield to principle explanation. After all, we now know that our time-honored relativity principle is precisely the principle that resolves the mystery of\cite{spooky} ``spooky actions at a distance.'' As Bell himself said\cite[p. 85]{bell1997} in 1990:
\begin{quote}
 I think the problems and puzzles we are dealing with here will be cleared up, and ... our descendants will look back on us with the same kind of superiority as we now are tempted to feel when we look at people in the late nineteenth century who worried about the aether. And Michelson-Morley ..., the puzzles seemed insoluble to them. And came Einstein in nineteen five, and now every schoolboy learns it and feels ... superior to those old guys. Now, it's my feeling that all this action at a distance and no action at a distance business will go the same way. But someone will come up with the answer, with a reasonable way of looking at these things. If we are lucky it will be to some big new development like the theory of relativity. 
\end{quote}
All of this is consistent with calls for a principle account of QM in general\cite{smolin2019,fuchs1,chiribella1,hardy2016,dakic,mullergroup}. Koberinski \&  M\"{u}ller write
\cite{koberinski2018},``By reconstructing the formalism of quantum theory in terms of operational constraints, one can cast quantum theory as a \textit{principle theory}, and thereby gain explanatory power regarding structural features of a quantum world.''

While our principle resolution of the trilemma of teaching the mystery of quantum entanglement does not constitute new physics as with SR, it is arguably ``some big new development'' pedagogically. Nonetheless, we don't expect to see it in the introductory physics textbooks anytime soon. Again, introductory physics textbooks should restrict their coverage to the most well-established material and the Topical Group of Quantum Information was only officially established as a Division of the American Physical Society in 2017. For example, questions on the Mathematical Tripos examination at Cambridge University contained reference to various ``jelly, froth, and vortex'' models\cite[pp. 236-240]{goldberg1984} of the aether until 1909. And, Maxwell published his famous equations in final form\cite{maxwell1865} in 1865, but Cunningham says the Tripos exam he took in 1902 barely covered electrodynamics since\cite[p. 239-240]{goldberg1984}, ``Maxwell's work was too recent and had not reached the textbook stage.'' In other words, faced again with historical precedent, we can expect there will be some serious delay between realization and widespread adoption of this principle account of quantum entanglement. In the meantime, it is up to each physics educator to decide what of this ``both new and true'' principle account to share with their students. After all, it is ultimately those students who will decide the fate of corresponding ``constructive efforts'' for resolving the mystery of quantum entanglement.


\section*{Appendix A}\label{SecAppendixA}

\noindent Here we briefly explain our class activity designed to introduce students to the mystery of entanglement per the Mermin device in a hands-on fashion. We break the class into groups of four -- Alice (A), Bob (B), Alice's particle (AP), and Bob's particle (BP). For class sizes that aren't divisible by four, the extra students are given the job of recording data. A and B are each given three folders numbered 1, 2, 3. AP and BP are each given six colored envelopes, three green (G) and three red (R). A and B sit opposite each other at a table and place their folders in a row in front of themselves on the table. The folder numbers will correspond to the setting numbers for the Mermin device, obviously. In each trial of the experiment, A and B close their eyes while AP randomly places one colored envelope in each of A's numbered folders and BP does the same with B's folders. In this first part of the activity, AP and BP are told to choose whatever color they want for each folder randomly and independently of each other. Having A and B close their eyes represents the fact that they can only know the color that resides in the one folder they choose to open in any given trial of the experiment. The fact that there are colors in all three folders for each trial even though A and B only actually open one folder is called \textit{counterfactual definiteness}. 

\clearpage

After AP and BP randomly place one colored envelope in each numbered folder, A and B open their eyes and randomly select one numbered folder to see what color envelope is inside. If, for example, A opens folder 2 and finds G while B opens folder 1 and finds R, the trial outcome is recorded as A and B choosing different settings and observing different colors. Overall, a count is kept for each possible trial outcome; SS is the total number of trials in which A and B choose the same settings and observe the same color, SD is the total number of trials with same settings and different colors, DS counts different settings and same colors, and DD counts counts different settings and different colors.  \textit{Realism} corresponds to the fact that A and B are only discovering what color envelope already exists in the folder that they choose to open, i.e., the act of opening the folder does not `create' the color inside. Counterfactual definiteness and realism are two assumptions that the Mermin device causes some physicists to question. 

\begin{table}
\begin{center}
\begin{tabular}{c|c|cc}
\multicolumn{1}{c}{} & \multicolumn{1}{c}{\bf Alice's and Bob's  } & \multicolumn{2}{c}{\bf Output Colors} \\
{\bf Model} & {\bf Settings (1, 2, or 3)} & Same  & Different  \\
\hline
\multirow{2}*{Purely Random} & Same  & $1/6$ & $1/6$ \\
& Different  & $1/3$ & $1/3$ \\
\hline
\multirow{2}*{Instruction Sets} & Same  & $1/3$ & $0$ \\
& Different  & $1/3$ & $1/3$ \\
\hline
\multirow{2}*{Quantum Mechanics} & Same  & $1/3$ & $0$ \\
& Different  & $1/6$ & $1/2$ \\
\end{tabular}
\caption{\textbf{Joint probability mass functions under all three models.}}
\label{tb:3}
\end{center}
\end{table}

The point of this part of the activity is to show that Fact 1 of the Mermin device is violated if we assume that AP and BP put the colored envelopes in the three folders uniformly at random. Recall that Fact 1 is that the observed colors are always the same when Alice and Bob choose the same settings (case (a)). It is helpful for an instructor to know how many trials are needed for students to see that the purely random model is inconsistent with Fact 1. The easiest way to check that inconsistency is to observe even a single instance of a different pair of colors when Alice's and Bob's settings are the same (i.e. $\mbox{SD} \geq 1$). From Table \ref{tb:3}, the probability of $\mbox{SD}=0$ in $N$ trials is $\left ( \frac 56\right )^N$, so $P(\mbox{SD} \geq 1) = 1 - \left ( \frac 56\right )^N \geq 0.95$ implies that $N\geq 17$. Thus, very few trials are needed to demonstrate that the ``purely random'' model of Table \ref{tb:3} is not consistent with QM.

The students are then shown Fact 1 of the Mermin device and asked how AP and BP would have to modify their behavior to guarantee Fact 1. They are told that AP and BP cannot know what folders A and B will select at the outset of any trial (this is called \textit{statistical independence}). They are also told that AP(BP) cannot base their decision as to what colors to place in which folders for A(B) based on the choices and outcomes for B(A) during the trial at hand (this assumption is called \textit{locality}). Statistical independence and locality are two more assumptions that the Mermin device causes some physicists to question. 

We have used this class activity for thousands of students since 1995 and in our experience it doesn't take the students long to realize that AP and BP need to use ``instruction sets'' per Section \ref{SecMerminDevice} to guarantee Fact 1 under the given constraints. The students are then told to repeat the experiment with AP and BP using randomly chosen instruction sets for each trial. The point here is to show that the instruction sets used to guarantee Fact 1 now violate Fact 2. That is, the data that the students produce using instruction sets should be statistically significantly different from the probabilities for QM in Table \ref{tb:3}.  

Again, it is useful for an instructor to know how many trials should be produced in the second part of the activity to confidently reject the ``instruction set'' model in Table \ref{tb:3}.  Using the hypotheses that the data are produced by a QM or an instruction set model, the Neyman-Pearson lemma implies that the most powerful hypothesis test is based on the statistic $ \mbox{DD} \ln (3/2) -\mbox{DS} \ln 2 $.  If the upper bounds on the probabilities of a Type I error and a Type II error are both $0.05$, then the number of trials needed is $N \geq 58$.  The mathematical details are omitted here to be concise, but they can readily be reproduced by any student who has had a course in mathematical statistics. Since we pool data across groups and a typical class has at least six groups, any one group doesn't need to do more than ten trials to convincingly show that instruction sets can't be used to explain how the Mermin device works and thereby establish the mystery in a hands-on fashion.



\clearpage

\section*{Appendix B}\label{SecAppendixB}

\noindent Since we have only shown how NPRF + $h$ reproduces Bell state entanglement for spin-$\frac{1}{2}$ particles and photons (as referenced), the reader may wonder about the generality of this principle account of entanglement. To see that NPRF + $h$ is indeed very general, we refer to the axiomatic reconstructions of quantum mechanics based on information-theoretic principles (the quantum reconstruction program, QRP). [For a detailed account of this Appendix see\cite{NPRF2024} ``Einstein's Entanglement: Bell Inequalities, Relativity, and the Qubit'' for the general reader or Stuckey, McDevitt, \& Silberstein\cite{stuckey2022} for a journal article.] There are many such information-theoretic derivations of the finite-dimensional Hilbert space formalism of QM with what M\"uller\cite{mueller2023} calls the ``first fully rigorous, complete reconstructions'' being produced by Chiribella, D’Ariano, \& Perinotti\cite{Chiribella2010} and Masanes \& M\"{u}ller\cite{masanesMuller2011} in 2010. For example, see those listed by Hardy\cite{hardy2016}, Dakic \& Brukner\cite{dakic2013}, and Jaeger\cite{jaeger2018}. 

\begin{figure}
\begin{center}
\includegraphics [height = 75mm]{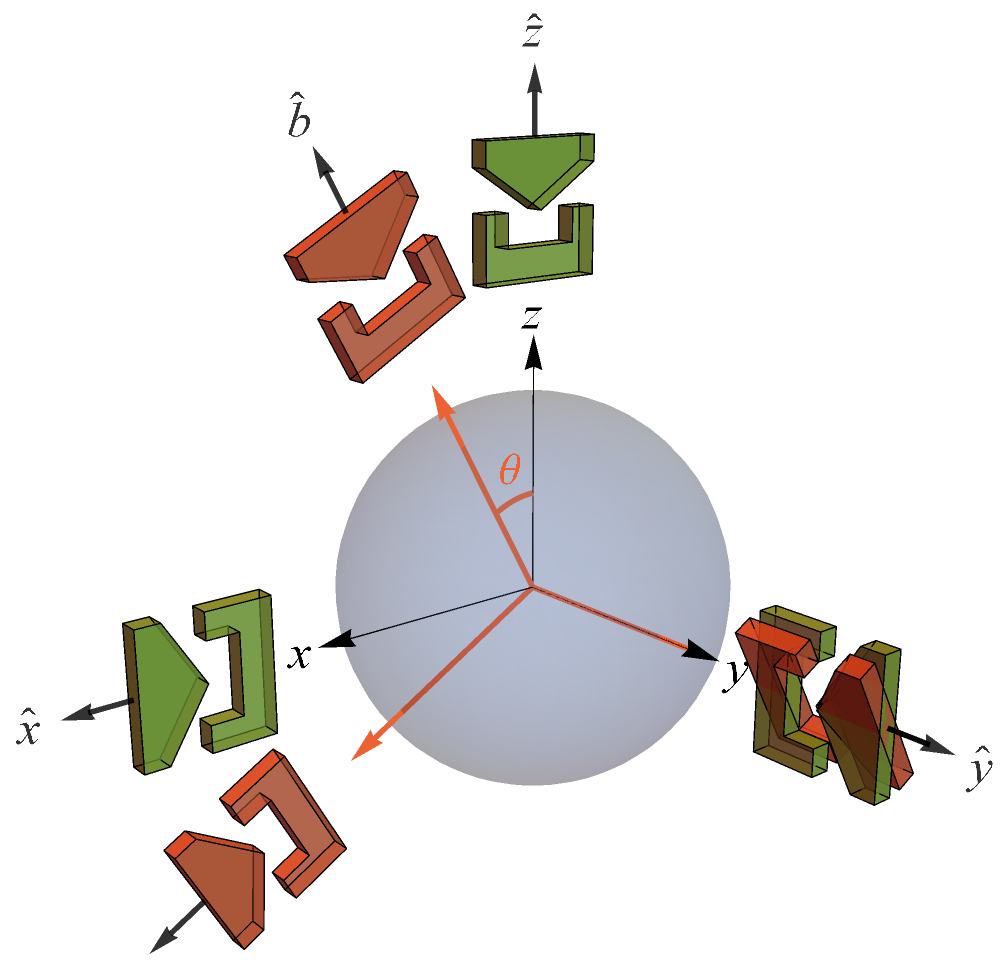} 
\caption{Two reference frames each associated with a set of mutually complementary SG spin measurements. Figure reproduced from Stuckey et al.\cite{stuckey2022} See also Brukner \& Zeilinger\cite{bruknerZeil2003}.} \label{ComplBases}
\end{center}
\end{figure}

In these reconstructions\cite{mullergroup}:
\begin{quote}
... at no point is it assumed that there are wave functions, operators or complex numbers – instead, those arise as consequences of the postulates. And we get all other ingredients and predictions of abstract finite-dimensional quantum: unitary transformations, uncertainty relations, the Schr\"{o}dinger equation (but not the choice of Hamiltonian or Lagrangian), Tsirelson’s bound on Bell correlations, and more.
\end{quote}
What these reconstructions show (one way or another) is that the (finite-dimensional) Hilbert space formalism of QM can be derived from an empirically discovered fact:
\begin{quote}
    Information Invariance \& Continuity: The total information of one bit is invariant under a continuous change between different complete sets of mutually complementary measurements.
\end{quote}
(Figure \ref{ComplBases}) due to Brukner \& Zeilinger\cite{brukner2009}, which is essentially the qubit. 

Starting with the qubit, $|\psi\rangle = c_1|u\rangle + c_2|d\rangle$ where $|c_1|^2 + |c_2|^2 = 1$, QRP builds the rest of the (finite-dimensional) Hilbert space structure of QM in composite fashion using their information-theoretic postulates. As Dakic \& Brukner point out\cite{dakic}, ``The fact that all other (higher-dimensional) systems can be built out of two-dimensional ones suggests that the latter can be considered as fundamental constituents of the world.'' Per Brukner \& Zeilinger\cite{bruknerZeil1999}, ``spin-$\frac{1}{2}$ affords a model of the quantum mechanics of all two-state systems, i.e. qubits,'' so what we concluded using SG spin measurements holds for qubits at the foundation of QM quite generally. 

That the fundamental constituent of the world is a qubit as opposed to some higher-dimensional bit ``can be shown with quite some effort'' using information-theoretic methods\cite{mullergroup}. Dakic \& Brukner write\cite{dakic2013}, ``we showed that the structure of the underlying probabilistic theory cannot be modified (for example by replacing quantum theory with a more general probabilistic theory) without changing the (three-) dimensionality of space.'' This is due to the fact that the measuring devices used to measure quantum systems are themselves made from quantum systems. For example, the classical magnetic field of an SG magnet is used to measure the spin of spin-$\frac{1}{2}$ particles and that classical magnetic field\cite{bruknergroup} ``can be seen as a limit of a large coherent state, where a large number of spin-$\frac{1}{2}$ particles are all prepared in the same quantum state.'' 

Given that the qubit forms the foundation of all (finite-dimensional) QM built in composite fashion, the most fundamental entangled states (upon which all others are built) are the Bell states, Eq. (\ref{BellStates}). As we showed, the mystery of entanglement per the Bell states resides in `average-only' conservation of whatever is represented by the $\pm 1$ measurement outcomes. And, again, that follows from the fact that the qubit is the fundamental constituent of the world and therefore cannot be subdivided, which we explained via NPRF applied to the rotational and translational invariance of these measurement outcomes in real space. Thus, Alice(Bob) says Bob(Alice) must average his(her) $\pm 1$ results to verify conservation of whatever is being measured, so we see that the mystery of `average-only' conservation is the result of conservation per NPRF. 

What makes the qubit different than a classical bit is that these different possible measurements are all related in continuous fashion. A classical bit has discrete measurement options, e.g., opening box 1 or box 2, each with two possible outcomes, e.g., find a ball or no ball. A qubit has continuous measurement options each with two possible outcomes. As Darrigol says\cite{darrigol2022}, the Hilbert space structure of QM:
\begin{quote}
    results from the harmonious blending of the discontinuity of measurement results with the continuity of the possibilities of measurement.
\end{quote}

\clearpage

\noindent For the SG spin measurements, the continuous measurements are the orientations of the SG magnets and the two outcomes are ``up'' or ``down'' relative to the N-S direction of those magnets. For polarizers, the continuous measurements are the orientations of the polarizing axis and the two outcomes are ``pass'' or ``no pass'' for the photons incident on the polarizer. For the double-slit experiment, the continuous measurements are locations of the detector screen relative to the slits along the optic axis and the two possible outcomes are ``slit 1'' or ``slit 2'' for a position measurement and ``constructive'' or ``destructive'' interference for a momentum measurement\cite{NPRF2024}. 

Per Goyal\cite{GoyalPhenomQBism2024}, the spatial notions of measurement in these examples are not inherent in the operational framework of QRP:
\begin{quote}
    For example, we tend to think of an agent as an embodied being localized in space; a physical system as an object that is spatially localized in our laboratory at all times; or a measurement as carried out by a chunk of equipment in one corner of a laboratory. But the operational framework abstracts away all of these \textit{spatial} notions. So, a \textit{physical system} is simply an entity that \textit{persists} -- it does not necessary exist anywhere in particular at a given moment in time. A \textit{measurement} is an abstract parameterized process that acts on a physical system to generate an \textit{outcome} and to output the same physical system -- it is not a spatially localized piece of equipment. The agent is simply \textit{an entity} that exists and persists over time, and is capable of \textit{observing outcomes} and of \textit{freely acting} to change settings associated with measurement and interaction devices -- it is not a spatially localized human being.
\end{quote}
In that sense, the Planck postulate may be viewed as a \textit{spatial interpretation} of Information Invariance \& Continuity, constituting the second step in Goyal's ``elucidative strategy'' for quantum theory. 

To complete the interpretation of Information Invariance \& Continuity, we note that each measurement is associated with a reference frame per its complementary measurements and these reference frames are related by spatial rotations or translations. The complementary spin measurements, e.g., $J_x$ and $J_z$, are related by spatial rotations (Figure \ref{ComplBases}) as are the complementary polarization measurements. In the double-slit experiment, the complementary measurement configurations of position and momentum are related by spatial translations\cite{NPRF2024}. 

Finally, all of this can be associated directly with Planck's constant $h$ since, as H\"ohn notes\cite{hohn2018}, $h$ represents ``a universal limit on how much simultaneous information is accessible to an observer.'' For example, for complementary spin measurements $(J_x,J_y,J_z)$ the commutator is $J_xJ_y - J_yJ_x=\textbf{i}\hbar J_z$ (which also applies to photon polarization\cite{Kim2010}). For the complementary measurements of position $x$ and momentum $p$ in the double-slit experiment the commutator is $xp - px = \textbf{i}\hbar$. And again, these complementary measurement configurations establish a reference frame related to other complementary measurement reference frames in continuous fashion via spatial rotations or translations. Therefore, the invariance of the total information between these different reference frames means $h$ is the same in reference frames related by spatial rotations and translations, i.e., we have the Planck postulate in analogy with the light postulate.

Putting all of this together we see that Information Invariance \& Continuity at the foundation of axiomatic reconstructions of QM is the information-theoretic counterpart to the conventional quantum characteristics of noncommutativity, superposition, and complementarity. And upon spatialization of QRP's operational notion of measurement, it entails the invariance of $h$ per the measurement outcomes in inertial reference frames of different complete sets of mutually complementary measurements, which can obviously be justified by the relativity principle (NPRF + $h$).

Therefore, NPRF + $h$ is fundamental to the finite-dimensional Hilbert space structure of QM precisely as NPRF + $c$ is fundamental to the Minkowski spacetime structure of SR. ``NPRF + $h$ + additional mathematical assumptions'' yields the Hilbert space of QM and ``NPRF + $c$ + additional mathematical assumptions'' yields the Lorentz transformations of SR. QM differs from classical physics precisely because $h$ is not zero and SR differs from Newtonian mechanics precisely because $c$ is not infinite. So, the physics educator can be assured that the analogy between the mysteries of time dilation and length contraction in SR and the mystery of quantum entanglement in QM as presented herein is indeed quite foundational.
\newpage

\end{document}